\documentclass[10pt,a4paper]{amsart}

\usepackage{amsmath,amsthm,amssymb,tikz}
\usepackage{mathtools}
\usepackage{enumitem}
\usepackage{hyperref}
\usepackage{orcidlink}

\newtheorem{theorem}{Theorem}[section]
\newtheorem{lemma}[theorem]{Lemma}
\newtheorem{proposition}[theorem]{Proposition}
\newtheorem{corollary}[theorem]{Corollary}
\theoremstyle{definition}
\newtheorem{definition}[theorem]{Definition}
\newtheorem{example}[theorem]{Example}
\newtheorem{remark}[theorem]{Remark}

\newcommand{\sem}[1]{[\![{#1}]\!]}
\newcommand{\diam}[1]{\langle {#1} \rangle}
\newcommand{\boxm}[1]{[{#1}]}
\newcommand{\norm}[1]{\lVert {#1} \rVert_\N}
\newcommand{{\Val}}{\mathsf{Val}}
\newcommand{{\Prop}}{\Phi_0}
\newcommand{{\Formu}}{\Phi}
\newcommand{{\AtProg}}{\Pi_0}
\newcommand{{\Prog}}{\Pi}
\newcommand{{\N}}{\mathbb{N}}
\newcommand{{\PDL}}{\mathtt{PDL}}
\newcommand{{\NPDL}}{\mathbb{N}\PDL}
\newcommand{{\FL}}{\textit{FL}}
\newcommand{{\Tau}}{\mathcal{T}}

\author[K. Manoorkar]{Krishna Manoorkar \orcidlink{0000-0003-3664-7757}}
\address{Institute of Computer Science of the Czech Academy of Sciences, Prague}
\email{manoorkar@cs.cas.cz \\ poiger@cs.cas.cz}
\author[W. Poiger]{Wolfgang Poiger \orcidlink{0009-0002-8485-5905}}

\title[Propositional Dynamic Logic with Program Costs]{Propositional Dynamic Logic with Program Costs}


\begin{document}
\begin{abstract}
We introduce $\NPDL$, a resource-aware variant of Propositional Dynamic Logic ($\PDL$) capable of accounting for program costs measured by natural numbers (such as time or space in the sense of computability theory). In this framework, programs correspond to transition systems weighted by extended natural numbers, and the syntax of our logic includes formulas expressing which properties are reachable by program executions within specific cost constraints. First, we explain how the standard program constructs of non-deterministic choice, sequential composition, iteration, and tests are interpreted in this weighted setting and establish the validity of their corresponding reduction axioms. Using an appropriate adaptation of the (Fischer-Ladner) filtration technique, we proceed to prove decidability and completeness for $\NPDL$ as our main results. We furthermore provide upper and lower complexity bounds for this logic using translations between standard $\PDL$ and $\NPDL$.  
\end{abstract}

\maketitle

\section{Introduction}
\label{sec:introduction}

In the landscape of formal methods in computer science, \emph{Propositional Dynamic Logic} ($\PDL$) is a well-known logical calculus for reasoning about programs to formally verify their correctness~\cite{FischerLadner1979,HarelKozen2000}. Combining modal logic and regular expressions, $\PDL$ specifies operational program constructs such as non-deterministic choice, sequential composition, iteration, and tests in a \emph{decidable} logical framework. More broadly, $\PDL$ can be conceived as a prototypical logic of `action and consequence'. For instance, this has been a primary inspiration for the `dynamic turn' in epistemic logic~\cite{vanDitmarsch2007}, where $\PDL$ and extensions thereof have been used to model concepts such as belief revision~\cite{vanEijck2008} or communicative updates~\cite{vanBenthem2006}.     
 
While standard $\PDL$ describes the qualitative reachability of states, in practice one may require additional \emph{quantitative} assessments. In the computational context, systems have critical constraints on resources such as time, memory, or energy. Therefore, it is reasonable to consider formal frameworks capable of reasoning about \emph{program efficiency}. Similar motivations arise in other areas, for instance due to the fact that actions of real-world agents are driven by considerations of \emph{cost and utility}.      

In this article we introduce $\NPDL$, a logic which includes discrete \emph{program costs} (e.g.\ time or space in the sense of computability  theory) into the $\PDL$ framework. Here, programs $\pi$ correspond to \emph{weighted} accessibility relations $R_\pi \colon X^2 \to \mathbb{N}^\infty$, where the weight $R_\pi(x,y)$ is an extended natural number modeling the (least) cost of reaching $y$ via executing $\pi$ at $x$. In particular, the \emph{addition} of natural numbers is seen as cost \emph{combination} and the weights $0$ and $\infty$ correspond to the usual accessibility and inaccessibility of $\PDL$. From this perspective, the \emph{weighted program constructs} such as composition and iteration we define for $\NPDL$ are chosen to naturally generalize those of standard $\PDL$ (Subsection~\ref{subsection:CostFrames}).

The syntax of $\NPDL$ includes formulas $\diam{\pi}^m\varphi$ for every (non-extended) natural number $m \in \mathbb{N}$, intuitively stating that `\emph{executing $\pi$ to reach $\varphi$ is possible for a cost of no more than $m$}'. That is, the `base logic' underlying $\NPDL$ is similar to weighted/chain-graded modal logic \cite{Larsen2014, Dong2021,Ma2015}. In the language of $\NPDL$, various cost-sensitive properties of programs are easily expressible, for instance $\varphi \rightarrow [\pi^\ast]^m\varphi$ can be read as `\emph{$\varphi$ is preserved while looping $\pi$ without exceeding cost $m$}'. Validities, for example of reduction axioms for weighted program constructs, can express program equivalences up to a cost (Subsection~\ref{subsection:SyntaxSemantics}).       

As the main results of this article, we prove \emph{decidability} (Section~\ref{section:Decidability}) and \emph{completeness} (Section~\ref{section:Completeness}) for $\NPDL$. Based on the above-mentioned reduction axioms, we define an adequate notion of \emph{Fischer-Ladner closure} $\FL(\varphi)$ for every formula $\varphi$. The size $|\FL(\varphi)|$ remains finite with upper bound $\norm{\varphi} \in \mathbb{N}$ parametric in $\varphi$, however this bound can be strictly larger than the length of $\varphi$ (Subsection~\ref{subsection:FL-closure}).
We then define appropriate (weighted) \emph{filtrations} and prove the corresponding Filtration Theorem, showing that $\NPDL$ has the \emph{bounded finite model property}
\footnote{Sometimes also called \emph{strong finite model property} \cite{BlackburnRijkeVenema2001} or \emph{small model property} \cite{HarelKozen2000}.}
(Subsection~\ref{subsection:FiltrationFMP}).  
To obtain decidability we also show that verifying satisfiability of $\varphi$ it suffices to restrict to models with \emph{bounded weights} at most $N(\varphi)+1 \in \mathbb{N}$, again parametric in $\varphi$ (Subsection~\ref{subsection:TowardsDecidability}). Using Fischer-Ladner filtrations of \emph{quasi-canonical models}, we prove that the reduction axioms of $\NPDL$ together with weighted generalizations of the induction rule yield a sound and (weakly) \emph{complete axiomatization} (Section~\ref{section:Completeness}). While our proofs are based on standard techniques (e.g.\ see \cite{HarelKozen2000}), the addition of costs presents additional technical challenges, which for instance becomes apparent in the proof of the Filtration Theorem (Theorems~\ref{theorem:FiltrationTheorem} \& \ref{theorem:FiltrationTheorem-II}), the novel use of `cut-off models' (Subsection~\ref{subsection:TowardsDecidability}), or the proof that the quasi-canonical model is standard with respect to sequential composition (Proposition~\ref{proposition:CanModel-Constructs}(2)). 

To obtain a \textsc{2exptime}-upper bound on the \emph{complexity} of $\NPDL$, we use a translation of $\NPDL$ into standard $\PDL$ over an extended set of atomic programs (Subsection~\ref{subsection:SatisfiabilityComplexityTranslations}). This translation is interesting in its own right, as it illustrates that $\NPDL$ is more succinct than standard $\PDL$. This is also apparent in the complexity of the model checking problem (Subsection~\ref{subsection:ModelCheckingComplexity}). The question whether $\NPDL$ remains \textsc{exptime}-complete is left as the main open problem. More details and further questions for future research (e.g.\ interpolation,  generalizations to other logics) are found in the concluding Section~\ref{section:Conclusion}.                  

\textbf{Related Work.} Other cost-sensitive versions of $\PDL$ have been proposed in the literature~\cite{Behounek2008, Sedlar2020,Sedlar2021}, however all of these are \emph{fuzzy or many-valued} (i.e.\ formulas can receive non-classical truth-values). While we retain the representation of costs via weighted accessibility relations used therein, note that in $\NPDL$ formulas remain Boolean-valued. In this sense, our setting more closely resembles that of \emph{graded-belief based programs} in~\cite{Laverny2005,Laverny2005b}, though these works are otherwise dissimilar in their motivation, set-up, and results. As mentioned above, the (non-dynamic) modal logic underlying $\NPDL$ is similar to \emph{weighted modal logic} \cite{Larsen2014, Dong2021} (which often uses real weights) and \emph{finite-chain graded modal logic} \cite{Ma2015} (note that this differs from the `graded modal logic' which counts successors). For instance, it is straightforward to adapt our filtration method to obtain decidability for the latter. Lastly, let us mention \cite{HansenPoiger2025}, which proves strong completeness of iteration-free chain-graded $\PDL$ as an instance of a \emph{coalgebraic dynamic logic} \cite{HansenKupke2015} (cf. Remark~\ref{remark:Coalgebra}).

\section{Propositional Dynamic Logic with Program Costs}
\label{section:PDLwithProgramCosts}
In this section, we explain how to interpret costs on frames weighted by extended natural numbers (Subsection~\ref{subsection:CostFrames}) and introduce $\PDL$ with program costs (Subsection~\ref{subsection:SyntaxSemantics}).
Throughout the paper, we assume that the reader is familiar with basic modal logic and standard propositional dynamic logic (e.g.\ see \cite{BlackburnRijkeVenema2001, HarelKozen2000}).
\subsection{Extended Natural Numbers and Cost Frames}
\label{subsection:CostFrames}
We exclusively deal with \emph{discrete} costs  modeled by \emph{extended natural numbers} $\N^\infty := \N \cup \{ \infty \}$, where the value $\infty$ is intuitively interpreted as an `unattainable' cost corresponding to \emph{inaccessibility}. We often think of the extended natural numbers as endowed with the \emph{tropical semiring} (respectively \emph{quantale}) structure  $\langle \N^\infty, \bigwedge, +, \infty, 0 \rangle$ 
where $\bigwedge$ is the minimum (which defines a complete lattice order, in particular $\bigwedge \varnothing = \infty$) and $+$ is addition of natural numbers with $\infty$ absorbing (i.e.\ $x + \infty = \infty + x = \infty$). We intuitively think of addition as cost \emph{combination} (other notions of cost combination can also be accommodated, see Remark~\ref{remark:OtherCombinations}).       

While semantics of standard modal logic (and $\PDL$) is based on Kripke frames (i.e.\ directed graphs), we use \emph{weighted} directed graphs for our underlying semantics.  

\begin{definition}[Cost frames]\label{defintion:CostFrames}
A \emph{cost frame} $\langle X, R \rangle$ consists of a set of \emph{states} $X$ together with a \emph{{weighted accessibility relation}} 
$R \colon X^2 \to \N^\infty$. 
\end{definition}

The value $R(x,y) \in \mathbb{N}^\infty$ models the (least) cost of the $R$-transition from $x$ to $y$. In particular, every Kripke frame $R \subseteq X^2$ can be naturally seen as a cost frame via $R(x,y) = 0 \Leftrightarrow xRy$ and $R(x,y) = \infty \Leftrightarrow  \neg x R y$. That is, cost $0$ represents standard accessibility (i.e.\ `for free') and cost $\infty$ represents standard inaccessibility (i.e.\ `never affordable'). This idea, albeit simple, guides many of our considerations in this article. For instance, the following generalizes the usual composition of binary relations in this sense. 
\begin{definition}[Sequential composition]\label{definition:Composition}
Let $\langle X,R_1 \rangle$ and $\langle X, R_2\rangle$ be cost frames. The \emph{sequential composition} of $R_1$ and $R_2$ is the weighted relation $R_1 \circ R_2 \colon X^2 \to \mathbb{N}^\infty$ defined by 
$$
(R_1 \circ R_2)(x,y) = \bigwedge_{v \in X} R_1(x,v) + R_2(v,y).
$$ 
\end{definition}
That is, $R_1 \circ R_2(x,y)$ represents the least cost to get from $x$ to $y$ by an $R_1$-transition followed by an $R_2$-transition (where $+$ corresponds to cost combination). 
Similarly, we generalize reflexive transitive closures. For any subset $S \subseteq X$, define the \emph{diagonal} $\Delta_S \colon X^2 \to \mathbb{N}^\infty$ by 
$$
\Delta_S(x,y) = \begin{cases}
0 & \text{ if } x = y \text{ \& } x \in S, \\
\infty & \text{ otherwise.} 
\end{cases}
$$   

\begin{definition}[Iteration]\label{definition:Iteration}
For $\langle X,R \rangle$ a cost frame, inductively define $R^{[k]}\colon X^2 \to \mathbb{N}^\infty$ for all $k \geq 0$ by $R^{[0]} = \Delta_X$ and $R^{[k+1]} = R \circ R^{[k]}$. The \emph{iteration} $R^\ast \colon X^2 \to \mathbb{N}^\infty$ is defined by 
$$
R^\ast(x,y) = \bigwedge_{k \geq 0} R^{[k]}(x,y).
$$   
\end{definition}

\begin{remark}[Other cost combinations]\label{remark:OtherCombinations}
Our definitions are based on the tropical semiring structure on $\mathbb{N}^\infty$, however one can develop similar frameworks for other quantales. For instance, one may consider $\langle \mathbb{N}^\infty, \bigwedge, \vee, \infty, 0 \rangle$, instead treating the \emph{maximum} $\vee$ as cost combination (e.g.\ , while $+$ as combination is reasonable for interpreting cost as \emph{time}, using $\vee$ as combination appears more reasonable for interpreting cost as \emph{space}). In this article, we focus on the (technically more challenging) case of $+$, however let us note that most of our results have analogues (oftentimes simpler) in the case of $\vee$.    
\end{remark}

\begin{remark}[Coalgebraic perspective \cite{HansenKupke2015, HansenPoiger2025}]\label{remark:Coalgebra}
Besides their cost interpretation, the above operations on cost frames arise naturally from a \emph{coalgebraic} perspective. That is, weighted cost frames can be identified with $\mathcal{P}_{\mathbb{N}^\infty}$-coalgebras where $\mathcal{P}_{\mathbb{N}^\infty} \colon X \mapsto (\mathbb{N}^\infty)^X$ is an appropriate generalization of the \emph{covariant powerset functor} $\mathcal{P}$ (whose coalgebras are standard Kripke frames). It can be checked (cf. \cite[Example 10(2)]{HansenPoiger2025}) that this functor is a \emph{monad}, the composition $\circ$ defined above is the corresponding \emph{Kleisli composition}, and  $(\cdot)^\ast$ can be seen as \emph{Kleisli iteration} (in particular noting that $\Delta$ relates to the unit of the monad $\mathcal{P}_{\mathbb{N}^\infty}$).   
\end{remark}

\subsection{Syntax and Semantics}
\label{subsection:SyntaxSemantics}

We now introduce $\NPDL$, our cost-sensitive variant of $\PDL$, in the following language. 

\begin{definition}[Formulas \& Programs]\label{definition:ProgramsFormulas}
We define the collections of \emph{formulas} $\varphi \in \Formu$ and \emph{programs} $\pi \in \Pi$ by simultaneous induction as follows. 
\begin{align*}
&&\varphi \ &::= \ p  
\text{ $\big\vert$ } \neg \varphi  
\text{ $\big\vert$ } \varphi_1 \wedge \varphi_2 \text{ $\big\vert$ } \diam{\pi}^m \varphi  
\\
&&\pi \ &::= \ a 
\text{ $\big\vert$ }  \pi_1 \cup \pi_2
\text{ $\big\vert$ } \pi_1 ; \pi_2 \text{ $\big\vert$ } \pi^\ast \text{ $\big\vert$ } \varphi?
\end{align*}
Here, $p \in \Prop$ and $a \in \AtProg$ range over countable collections of \emph{propositional variables} and \emph{atomic programs}, respectively, and $m \in \mathbb{N}$ ranges over the (non-extended) natural numbers. 
\end{definition}

We define additional propositional connectives such as $\vee, \rightarrow, \top, \bot$ as usual and we use $\boxm{\pi}^m \varphi := \neg \diam{\pi}^m \neg \varphi$. As in standard $\PDL$ we refer to the program constructs $\cup$ as \emph{non-deterministic choice}, $;$ as \emph{sequential composition}, and $(\cdot)^\ast$ as \emph{iteration}. Programs of the form $\psi?$ are called \emph{tests}. For our semantics we assign cost frames (Definition~\ref{defintion:CostFrames}) to programs.  

\begin{definition}[Models]
\label{definition:Model}
A \emph{model} is a structure $\mathfrak{M} = \langle X, \{ R_a \}_{a \in \AtProg}, \Val \rangle$ such that $\langle X, R_a\rangle$ is a cost frame for every atomic program $a \in \AtProg$ and $\Val \colon \Prop \to \mathcal{P}(X)$ is a propositional valuation. 
\end{definition}

\begin{definition}[Satisfaction]
\label{definition:Satisfaction}
For a model $\mathfrak{M} = \langle  X, \{ R_a \}_{a \in \AtProg}, \Val \rangle$ and $x \in X$, inductively define $\mathfrak{M}, x \Vdash \varphi$ for all $\varphi \in \Formu$ and $R_\pi \colon X^2 \to \N^\infty$ for all $\pi \in \Prog$: \\
\textbf{\textit{Formulas}:}
\begin{itemize} 
\item $\mathfrak{M}, x \Vdash p  \phantom{i}\text{ iff }\phantom{i}   x \in \Val(p)$
\item $\mathfrak{M}, x \Vdash \varphi \wedge \psi \phantom{i}\text{ iff }\phantom{i} \mathfrak{M}, x \Vdash \varphi \And \mathfrak{M}, x \Vdash \psi$
\item $\mathfrak{M}, x \Vdash \neg\varphi  \phantom{i}\text{ iff }\phantom{i} \mathfrak{M}, x \not\Vdash \varphi$
\item $\mathfrak{M}, x \Vdash \diam{\pi}^m \varphi \phantom{i}\text{iff}\phantom{i} \exists y\colon R_\pi(x, y) \leq m \And \mathfrak{M}, y \Vdash \varphi$
\end{itemize}
\textbf{\textit{Programs}:}
\begin{itemize}
\item $R_{\pi_1 \cup \pi_2} \phantom{i}=\phantom{i} R_{\pi_1} \wedge R_{\pi_2}$ 
\item $R_{\pi_1;\pi_2} \phantom{i}=\phantom{i} R_{\pi_1} \circ R_{\pi_2}$  
\item $R_{\pi^\ast} \phantom{i}=\phantom{i}  (R_\pi)^\ast$
\item $R_{\psi?} \phantom{i} = \phantom{i} \Delta_{\sem{\psi}} \phantom{i}$ $(\sem{\psi} = \{ x \mid \mathfrak{M}, x \Vdash \psi \})$
\end{itemize}
($R_{1} \wedge R_{2}$ is the point-wise minimum. For $\circ$, $\Delta$, and $R^\ast$ recall Definitions~\ref{definition:Composition}$\And$\ref{definition:Iteration}.)   
\end{definition}

As usual, we say that $\varphi$ is \emph{satisfiable} if $\mathfrak{M},x \Vdash \varphi$ holds at some state of some model and \emph{valid} if $\mathfrak{M}, x \Vdash \varphi$ holds at all states of all models.

By definition, the clause for the dual modalities is $\mathfrak{M}, x \Vdash \boxm{\pi}^m\varphi$ iff $\forall y\colon R_{\pi}(x,y) \leq m  \ \Rightarrow \ \mathfrak{M}, y \Vdash \varphi$.
The formula $\diam{\pi}^m\varphi$ intuitively states that `\emph{executing $\pi$ to reach a state where $\varphi$ holds is possible for a cost of no more than $m$}' and $\boxm{\pi}^m \varphi$ expresses that `\emph{every execution of $\pi$ on a resource-budget at most $m$ results in property $\varphi$}'. With this, further cost-sensitive properties about programs can be expressed, for instance $\diam{\pi}^m \top$ can be read as `\emph{program $\pi$ terminates without costing more than $m$}' and $\varphi \rightarrow [\pi^\ast]^m\varphi$ can be read as `\emph{$\varphi$ is preserved when looping $\pi$ as long as cost $m$ is not exceeded}'. 

As mentioned before, the weights on the accessibility relation $R_\pi$ correspond to the \emph{(least) costs of executing program $\pi$}. This interpretation also pertains to more complex programs, for instance the program $\mathtt{if } \phantom{i} \varphi \phantom{i} \mathtt{ then } \phantom{i} \pi_1 \phantom{i}\mathtt{ else } \phantom{i}\pi_2 := \varphi? ; \pi_1 \cup \neg \varphi?; \pi_2$ corresponds to 
$$
R_{\mathtt{if } \phantom{i} \varphi \phantom{i} \mathtt{ then } \phantom{i} \pi_1 \phantom{i}\mathtt{ else } \phantom{i}\pi_2}(x,y) \phantom{i}=\phantom{i} \begin{cases}
R_{\pi_1}(x,y) & \text{if } \mathfrak{M}, x \Vdash \varphi, \\
R_{\pi_2}(x,y) & \text{otherwise.}
\end{cases}
$$

Note that we implicitly assume (by definition of $\Delta$) that testing itself does not have a cost. However, such costs can also be accommodated, e.g.\ using `indexed tests' $R_{\psi?^n} = \Delta_{\sem{\psi}} \vee n$.

In the following two examples, we consider some sample applications of $\NPDL$.

\begin{example}[Resource-bounded verification of a countdown loop]
\label{example:CountdownLoop}
Consider a simple program that repeatedly decrements a non-negative
integer counter until it reaches zero. We use a cost model in which
each decrement operation has cost $1$, while evaluating a test has
cost $0$. This allows us to interpret the cost of an execution as the
number of decrement operations performed.

We model the program configurations by the set of states
$X = \mathbb{N}$, where each state $n$ represents a configuration in
which the counter has value $n$. Let $\mathsf{positive}$ and
$\mathsf{zero}$ be propositional variables with valuations
\[
  \Val(\mathsf{positive}) = \mathbb{N}{\setminus}\{0\},
  \qquad
  \Val(\mathsf{zero}) = \{0\}.
\]
Furthermore, let $\mathsf{dec}$ be an atomic program representing the
decrement operation. Its cost relation is defined by
\[
  R_{\mathsf{dec}}(n,m) =
  \begin{cases}
    1 & \text{if } n>0 \text{ and } m=n-1,\\
    \infty & \text{otherwise}.
  \end{cases}
\]
Thus, decrementing a positive counter incurs cost $1$, whereas no
decrement transition is available from the state $0$. The body of the loop and the entire loop can now be represented by
the program
\[
  \pi_{\mathsf{dec}} := (\mathsf{positive}?\,;\,\mathsf{dec})^\ast.
\]
The test ensures that the counter is positive before each
decrement. Consequently, every iteration of $\delta$ is executable
at cost $1$, and the loop terminates when the counter reaches zero.

Let us first consider the cost relation induced by the loop.
Starting in state $n$, executing exactly $k$ iterations, where
$0\leq k\leq n$, reaches state $n-k$ at cost $k$. No execution can
perform more than $n$ iterations. It follows that
\[
  R_{\pi_{\mathsf{dec}}}(n,m) =
  \begin{cases}
    n-m & \text{if } 0\leq m\leq n,\\
    \infty & \text{otherwise}.
  \end{cases}
\]
In particular, the least cost of reaching the zero state is $ R_{\pi_{\mathsf{dec}}}(n,0)=n$.
This observation can be expressed directly in $\NPDL$. For every
initial counter value $n\in\mathbb{N}$ and every cost bound
$B\in\mathbb{N}$, we have
\[
  \mathfrak{M},n \Vdash
  \diam{\pi_{\mathsf{dec}}}^{B}\mathsf{zero}
  \quad\Longleftrightarrow\quad n\leq B.
\]
Hence, the formula
$\diam{\pi_{\mathsf{dec}}}^{B}\mathsf{zero}$ expresses that the
countdown loop can reach a configuration in which the counter is
zero within the available cost budget $B$. In this example, the
formula holds precisely when the budget is sufficient to complete
the loop.

For instance, if the initial counter value is $3$, the execution is
\[
  3 \xrightarrow{1} 2
    \xrightarrow{1} 1
    \xrightarrow{1} 0,
\]
where each transition is labelled by its cost. The formula
$\diam{\pi_{\mathsf{dec}}}^{2}\mathsf{zero}$ is false at state $3$,
whereas $\diam{\pi_{\mathsf{dec}}}^{3}\mathsf{zero}$ is true there.
The former budget is insufficient to reach the zero state, while the
latter is sufficient.

The example also illustrates the role of the dual modality. Suppose
that $\mathsf{safe}$ is a propositional variable describing the
configurations that satisfy a given safety requirement. Then $\boxm{\pi_{\mathsf{dec}}}^{B}\mathsf{safe}$ expresses that every state reachable from the initial configuration
with least execution cost at most $B$ satisfies the safety
requirement. Since the program uses iteration, this includes states
reached after any number of loop iterations, including the initial
state itself.

Finally, the cost assignment can be modified to model other resource
measures. For example, if every decrement operation has cost $c>0$
instead of cost $1$, the least cost of reaching zero from state $n$
becomes $cn$, and the termination formula holds precisely when
$cn\leq B$. More generally, assigning different costs to different
decrement operations allows the logic to express termination
requirements under non-uniform resource consumption.
\hfill $\blacksquare$
\end{example}

\begin{example}[Budgeted network routing]
\label{example:NetworkRouting}
Consider a communication network whose nodes are states of a model
$\mathfrak{M}$. Let $\mathsf{forward}$ be an atomic program representing
the forwarding of a packet between adjacent nodes, where the cost
$R_{\mathsf{forward}}(x,y)$ represents the latency or energy required
to forward a packet from $x$ to $y$. A cost of $\infty$ indicates
that no direct transition is available.

Suppose the network contains four nodes $s,u,v,t$ and the following
transitions:
\[
\begin{array}{c|c}
\text{Transition} & \text{Cost}\\ \hline
s\to u & 1\\
u\to t & 4\\
s\to v & 2\\
v\to t & 1
\end{array}
\]
All other transitions have cost $\infty$. Let
$\mathsf{delivered}$ be a propositional variable that holds precisely
at the destination node $t$.

There are two routes from $s$ to $t$: the route through $u$ has
total cost $1+4=5$, whereas the route through $v$ has total cost
$2+1=3$. Consequently,
\[
R_{\mathsf{forward}^{\ast}}(s,t)
=\min\{1+4,\,2+1\}=3.
\]
The formula
\[
\diam{\mathsf{forward}^{\ast}}^{B}\mathsf{delivered}
\]
therefore holds at $s$ if and only if $B\geq 3$. In particular,
$\diam{\mathsf{forward}^{\ast}}^{2}\mathsf{delivered}$ is false at
$s$, whereas
$\diam{\mathsf{forward}^{\ast}}^{3}\mathsf{delivered}$ is true.

Thus, $\NPDL$ can express whether a destination is reachable within
a prescribed resource budget, while the semantics of sequential
composition and iteration computes the least cost over all possible
routes. Depending on the interpretation of transition costs, the
same model can represent latency-constrained routing, energy-aware
communication, or other budget-constrained reachability problems.
\hfill $\blacksquare$
\end{example}

We end this section with some validities of $\NPDL$. Note that \emph{(4)}$\And$\emph{(7)} for choice and test below are the same as in $\PDL$, while \emph{(5)}$\And$\emph{(6)} for composition and iteration are more involved.         

\begin{proposition}\label{proposition:ReductionAxiomsValid}
The following are valid $\NPDL$-formulas. 
\begin{enumerate}
\item $\phantom{a}\diam{\pi}^m \bot \phantom{i}\leftrightarrow\phantom{i} \bot$
\item $\phantom{a}\diam{\pi}^m (\varphi \vee \psi) \phantom{i}\leftrightarrow\phantom{i} \diam{\pi}^m \varphi \vee \diam{\pi}^m \psi$
\item $\phantom{a}\diam{\pi}^n \varphi \phantom{i}\rightarrow\phantom{i} \diam{\pi}^m \varphi \phantom{i}$ for all $n \leq m$
\item $\phantom{a}\diam{\pi_1 \cup \pi_1}^m \varphi \phantom{i}\leftrightarrow\phantom{i} \diam{\pi_1}^m \varphi \vee \diam{\pi_2}^m \varphi$
\item $\phantom{a}\diam{\pi_1 ; \pi_2}^m \varphi \phantom{i}\leftrightarrow\phantom{i} \bigvee\limits_{i = 0}^m \diam{\pi_1}^{m-i} \diam{\pi_2}^i \varphi$
\item $\phantom{a}\diam{\pi^\ast}^m \varphi \phantom{i}\leftrightarrow\phantom{i} \varphi \vee \bigvee\limits_{i = 0}^{m} \diam{\pi}^{m-i} \diam{\pi^\ast}^{i}\varphi $
\item $\phantom{a}\diam{\psi?}^m \varphi \phantom{i}\leftrightarrow\phantom{i} \psi \wedge \varphi$
\end{enumerate}
\end{proposition}
\begin{proof}
Let $\mathfrak{M} = \langle  X, \{ R_a \}_{a \in \AtProg}, \Val \rangle$ be a model.
The proofs of \emph{(1)}--\emph{(3)} are completely straightforward using the satisfiability clause for $\mathfrak{M}, x \Vdash \diam{\pi}^m\varphi$ of Definition~\ref{definition:Satisfaction}. \\
\emph{(4)}: Follows from $R_{\pi_1}(x,y) \wedge R_{\pi_2}(x,y) \leq m$ if and only if $R_{\pi_1}(x,y) \leq m$ or $R_{\pi_2}(x,y) \leq m$. \\
\emph{(5)}: $\mathfrak{M}, x \Vdash \diam{\pi_1;\pi_2}^m\varphi$ iff there exist $v, y \in X$ such that $R_{\pi_1}(x,v) + R_{\pi_2}(v,y) \leq m$ and $\mathfrak{M}, y \Vdash \varphi$. The equivalence to the right-hand side of the formula is clear with $i := R_{\pi_2}(v,y)$. \\
\emph{(6)}: $\mathfrak{M}, x \Vdash \diam{\pi^\ast} \varphi$ iff there is $y \in X$ and $k \geq 0$ such that $(R_\pi)^{[k]}(x,y) \leq m$ and $\mathfrak{M}, y \Vdash \varphi$. The case $(R_{\pi})^{[0]}(x,y) \leq m$ is equivalent to $x = y$ and $\mathfrak{M}, x \Vdash \varphi$. The case of $(R_{\pi})^{[k+1]}(x,y)$ with $k \geq 0$ holds if and only if there is $v \in X$ such that $R_\pi(x,v) + (R_{\pi})^{[k]}(v,y) \leq m$. Setting $i := (R_{\pi})^{[k]}(v,y)$ this is equivalent to $\mathfrak{M}, x \Vdash \diam{\pi}^{m-i}\diam{\pi^\ast}^i \varphi$. \\
\emph{(7)}: $\mathfrak{M}, x \Vdash \diam{\psi?}^m \varphi$ iff there is $y \in X$ with $R_{\psi?}(x,y) \leq m$ and $\mathfrak{M}, y \Vdash \varphi$, and the former is equivalent to $x = y$ and $\mathfrak{M}, x \Vdash \psi$.
\end{proof}

These validities motivate our definition of Fischer-Ladner closure (Definition~\ref{definition:FL-closure}) and are used in the proof of the corresponding Filtration Theorem (Theorem~\ref{theorem:FiltrationTheorem}) in the next section.

\section{Decidability}\label{section:Decidability}
In this section, we prove that $\NPDL$ is decidable, which we do in two steps. First, we show the \emph{bounded finite model property} using filtrations through the \emph{Fischer-Ladner closure} adequately defined for our framework (Subsections~\ref{subsection:FL-closure} \& \ref{subsection:FiltrationFMP}). Second, we show that there is a \emph{bound on weights} of atomic programs sufficient to check satisfiability of a formula (Subsection~\ref{subsection:TowardsDecidability}).
\subsection{The Fischer-Ladner Closure}\label{subsection:FL-closure}
We aim to prove the bounded finite model property using filtrations through an appropriately defined \emph{Fischer-Ladner closure} \cite[Theorem 3.2]{FischerLadner1979}, a finite set of formulas closed under reduction instances for program constructs. Our presentation here is similar to \cite[Section 6]{HarelKozen2000}.

\begin{definition}[Fischer-Ladner closure]\label{definition:FL-closure}
For all formulas $\varphi$ and programs $\pi$ we define the sets of formulas $\FL(\varphi)$ and $\FL^\Diamond\big(\diam{\pi}^m \varphi\big)$ for all $m \in \mathbb{N}$ as follows. \\
\textbf{\textit{Formulas}:}
\begin{itemize}
\item $\FL(p) = \{ p \}$ for propositional variables $p \in \Prop$,  
\item $\FL(\neg \varphi) = \{ \neg \varphi \} \cup \FL(\varphi)$, 
\item $\FL(\varphi_1 \wedge \varphi_2) = \{ \varphi_1 \wedge \varphi_2 \} \cup \FL(\varphi_1) \cup \FL(\varphi_2)$, 
\item $\FL\big(\diam{\pi}^m \varphi\big) = \FL^\Diamond\big(\diam{\pi}^m \varphi\big) \cup \FL(\varphi)$,
\end{itemize}
\textbf{\textit{Programs}:}
\begin{itemize}
\item $\FL^\Diamond\big(\diam{a}^m \varphi\big) = \{ \diam{a}^m \varphi \}$ for atomic programs $a \in \AtProg$, 
\item $\FL^\Diamond\big(\diam{\pi_1 \cup \pi_2}^m \varphi\big) = \{ \diam{\pi_1 \cup \pi_2}^m \varphi \} \cup  \FL^\Diamond\big(\diam{\pi_1}^m\varphi\big) \cup \FL^\Diamond\big(\diam{\pi_2}^m\varphi\big)$,
\item $\FL^\Diamond\big(\diam{\pi_1;\pi_2}^m \varphi\big) = \{ \diam{\pi_1;\pi_2}^m \varphi \} \cup {\bigcup\limits_{i = 0}^{m}\FL^\Diamond\big(\diam{\pi_1}^{m-i}\diam{\pi_2}^{i} \varphi\big) \cup \FL^\Diamond \big(\diam{\pi_2}^{i} \varphi\big)}$, 
\item $\FL^\Diamond\big(\diam{\pi^\ast}^m\varphi\big) = \bigcup\limits_{n+n' \leq m} \FL^\Diamond\big(\diam{\pi}^{n}\diam{\pi^\ast}^{n'} \varphi\big) \cup \{ \diam{\pi^\ast}^{n'}\varphi \}$, 
\item $\FL^\Diamond\big(\diam{\psi?}^m\varphi\big) = \{ \diam{\psi?}^m\varphi \} \cup \FL(\psi)$.
\end{itemize} 
We refer to the collection of formulas $\FL(\varphi)$ as the \emph{Fischer-Ladner closure} of $\varphi$.
\end{definition}

Our first goal is to show that $\FL(\varphi)$ is always \emph{finite} with an upper bound $\norm{\varphi}$ parametric in $\varphi$. Note that, due to the closure conditions for sequential composition and iteration above, this upper bound can \emph{not} simply be the length of $\varphi$ (as is the case in standard $\PDL$). 

\begin{definition}\label{definition:NormPhi}
For all formulas $\varphi$ and programs $\pi$ we define the natural numbers $\norm{\varphi}$ and $\norm{\diam{\pi}^m}$ for every $m \in \mathbb{N}$ as follows.
\\
\textbf{\textit{Formulas}:}
\begin{itemize}
\item $\norm{p} = 1$ for propositional variables,  
\item $\norm{\neg \varphi} = 1 + \norm{\varphi}$,  
\item $\norm{\varphi_1 \wedge \varphi_2} = 1 + \norm{\varphi_1} + \norm{\varphi_2}$, 
\item $\norm{\diam{\pi}^m \varphi} = \norm{\diam{\pi}^m} + \norm{\varphi}$.
\end{itemize}
\textbf{\textit{Programs}:}
\begin{itemize}
\item $\norm{\diam{a}^m} = 1$ for atomic programs, 
\item $\norm{\diam{\pi_1 \cup \pi_2}^m} = 1 + \norm{\diam{\pi_1}^m} + \norm{\diam{\pi_2}^m}$, 
\item $\norm{\diam{\pi_1;\pi_2}^m} = 1 + \sum\limits_{i = 0}^m \norm{\diam{\pi_1}^{m-i}} + \norm{\diam{\pi_2}^{i}}$,
\item $\norm{\diam{\pi^\ast}^m} = (m+1) + \sum\limits_{n + n' \leq m} \norm{\diam{\pi}^n}$,
\item $\norm{\diam{\psi?}^m} = 1 + \norm{\psi}$.
\end{itemize}
(Since only finite sums are involved, all of these are well-defined natural numbers.) 
\end{definition}

\begin{lemma}\label{lemma:FL-Finite}
For all formulas $\varphi$ it holds that $|\FL(\varphi)| \leq \norm{\varphi}$.
\end{lemma}
\begin{proof}
By simultaneous induction show for all formulas $\varphi$ and programs $\pi$ that
$$
|\FL(\varphi)| \phantom{i} \leq \phantom{i} \norm{\varphi} \quad \text{ and } \quad |\FL^\Diamond\big(\diam{\pi}^m \varphi \big)|\phantom{i} \leq \phantom{i}\norm{\diam{\pi}^m}.
$$ 
The formula induction is straightforward using the inductive definition of $\FL(\varphi)$ in Definition~\ref{definition:FL-closure} and the corresponding inductive hypotheses. For sequential composition, we compute
{\small
\begin{align*}
|\FL^\Diamond\big(\diam{\pi_1 ; \pi_2}^m \varphi\big)| \phantom{a} & \leq \phantom{a} 1 \phantom{i}+\phantom{i} \sum_{i = 0}^m |\FL^\Diamond\big(\diam{\pi_1}^{m-i}\diam{\pi_2}^{i}\varphi\big)| + |\FL^\Diamond\big(\diam{\pi_2}^{i}\varphi\big)| \\ 
&  \leq \phantom{a} 1 \phantom{i}+\phantom{i} \sum_{i = 0}^m \norm{\diam{\pi_1}^i} + \norm{\diam{\pi_2}^{m-i}} \phantom{a} = \phantom{a} \norm{\diam{\pi_1; \pi_2}^m}
\end{align*}
}
where the second inequality uses the inductive hypotheses on $\pi_1$ and $\pi_2$.

For iteration, first we note that as a direct consequence of Definition~\ref{definition:FL-closure}, it holds that 
$
\FL^\Diamond(\diam{\pi^\ast}^m\varphi) \subseteq \{ \diam{\pi^\ast}^i \varphi \mid 0 \leq i \leq m \} \cup \bigcup \{ \FL^\Diamond(\diam{\pi}^{n}\diam{\pi^\ast}^{n'} \varphi) \mid n + n' \leq m\}
$, and therefore 
{\small
\begin{align*}
|\FL^\Diamond\big(\diam{\pi^\ast}^m\varphi\big)| & \phantom{a} \leq \phantom{a} |\{ \diam{\pi^\ast}^i \varphi \mid 0 \leq i \leq m \}| \phantom{i} + \sum_{n + n' \leq m} | \FL^\Diamond\big(\diam{\pi}^{n}\diam{\pi^\ast}^{n'} \varphi\big)| \\ 
& \phantom{a} \leq \phantom{a} (m+1) \phantom{i} + \sum_{n + n' \leq m} \norm{\diam{\pi}^n} \phantom{a} = \phantom{a} \norm{\diam{\pi^\ast}^m}
\end{align*} 
}
as desired, where the second inequality uses the inductive hypothesis on $\pi$.
\end{proof}

To prove the Filtration Theorem (Theorem~\ref{theorem:FiltrationTheorem}), we use various closure properties of $\FL(\varphi)$ (Corollary~\ref{corollary:FL-Closure-Rules}), all of which follow from the next lemma (similar to \cite[Lemma 6.1]{HarelKozen2000}).   

\begin{lemma}\label{lemma:FL-Closure-Rules}
For all formulas $\varphi$ and programs $\pi$ the following hold.
\begin{enumerate}[label = (\roman*)]
\item If $\psi \in \FL(\varphi)$ then $\FL(\psi) \subseteq \FL(\varphi)$,
\item If $\psi \in \FL^\Diamond\big(\diam{\pi}^m\varphi\big)$ then $\FL(\psi) \subseteq \FL\big(\diam{\pi}^m \varphi\big)$.
\end{enumerate} 
\end{lemma}
\begin{proof}
By simultaneous induction we show \emph{(i)} for all formulas $\varphi$ and \emph{(ii)} for all programs $\pi$. 

\textbf{Formula induction.} The case $p \in \Prop$ is due to $\FL(p) = \{ p \}$ and the cases $\neg\varphi$ and $\varphi_1 \wedge \varphi_2$ are clear by inductive hypotheses and definitions of $\FL(\neg\varphi)$ and $\FL(\varphi_1 \wedge \varphi_2)$. 

\emph{$\bullet$ Case $\diam{\pi}^m \varphi$}:
If $\psi \in \FL\big(\diam{\pi}^m \varphi\big)$, then by definition $\psi \in \FL(\varphi)$ or $\psi \in \FL^\Diamond\big(\diam{\pi}^m\varphi\big)$. In the former case use the inductive hypothesis \emph{(i)} to find $\FL(\psi) \subseteq \FL(\varphi) \subseteq \FL\big(\diam{\pi}^m \varphi\big)$ and in the latter case use the inductive hypothesis \emph{(ii)} to find $\FL(\psi) \subseteq \FL\big(\diam{\pi}^m \varphi\big)$.   

\textbf{Program induction.} The atomic case $a \in \AtProg$ is clear by $\FL^\Diamond\big(\diam{a}^m \varphi\big) = \{ \diam{a}^m \varphi \}$.  

\emph{$\bullet$ Case $\pi_1 \cup \pi_2$} : If $\psi \in \FL^\Diamond\big(\diam{\pi_1 \cup \pi_2}^m\varphi\big)$, then $\psi = \diam{\pi_1 \cup \pi_2}^m\varphi$ or $\psi \in \FL^\Diamond\big(\diam{\pi_1}^m \varphi\big)$ or $\psi \in \FL^\Diamond\big(\diam{\pi_2}^m \varphi\big)$. In the first case the statement is clear and in the other two cases the inductive hypothesis on $\pi_j \in \{ \pi_1, \pi_2 \}$ yields $\FL(\psi) \subseteq \FL\big(\diam{\pi_j}^m \varphi\big) \subseteq \FL\big(\diam{\pi_1 \cup \pi_2}^m \varphi\big)$.   

\emph{$\bullet$ Case $\pi_1 ; \pi_2$} : Suppose $\psi \in \FL^\Diamond\big(\diam{\pi_1 ; \pi_2}^m\varphi\big)$ and $\psi \neq \diam{\pi_1 ; \pi_2}^m\varphi$ (otherwise the statement is clear). Then there is some $i \in \{ 0, \dots, m \}$ such that $\psi \in \FL^\Diamond\big(\diam{\pi_1}^{m-i} \diam{\pi_2}^{i}\varphi\big)$ or $\psi \in \FL^\Diamond\big(\diam{\pi_2}^{i}\varphi\big)$. In the former case use the inductive hypothesis on $\pi_1$ to obtain $\FL(\psi) \subseteq \FL\big(\diam{\pi_1}^{m-i} \diam{\pi_2}^{i} \varphi\big) \subseteq \FL\big(\diam{\pi_1; \pi_2}^m \varphi\big)$ and in the latter case use the inductive hypothesis on $\pi_2$ to find $\FL(\psi) \subseteq \FL\big(\diam{\pi_2}^{i}\varphi\big) \subseteq \FL\big(\diam{\pi_1 ; \pi_2}^m \varphi\big)$.  

\emph{$\bullet$ Case $\pi^\ast$} : First note that by definition $n' \leq m$ implies $\FL^\Diamond\big(\diam{\pi^\ast}^{n'}\varphi\big) \subseteq \FL^\Diamond\big(\diam{\pi^\ast}^{m}\varphi\big)$. If $\psi \in \FL^\Diamond\big(\diam{\pi^\ast}^m \varphi\big)$, then there are $n + n' \leq m$ with $\psi = \diam{\pi^\ast}^{n'} \varphi$ or $\psi \in \FL^\Diamond\big(\diam{\pi}^n\diam{\pi^\ast}^{n'} \varphi\big)$. In the former case $\FL(\psi) \subseteq \FL\big(\diam{\pi^\ast}^m \varphi\big)$ follows from the above observation and in the latter case use the inductive hypothesis on $\pi$ to find $\FL(\psi) \subseteq \FL\big(\diam{\pi}^n \diam{\pi^\ast}^{n'} \varphi\big) \subseteq \FL\big(\diam{\pi^\ast}^m \varphi\big)$, where the last inclusion again uses the above observation.  

\emph{$\bullet$ Case $\rho?$} : If $\psi \in \FL^\Diamond\big(\diam{\rho?}^m \varphi\big)$, then $\psi = \diam{\rho}^m \varphi$ or $\psi \in \FL(\rho)$. The former case is clear, in the latter case inductive hypothesis \emph{(i)} on $\rho$ yields $\FL(\psi) \subseteq \FL(\rho) \subseteq \FL\big(\diam{\rho?}^m \varphi\big)$.
\end{proof}

With this, we get the following crucial closure rules of the Fischer-Ladner closure with respect to the program constructs of $\NPDL$. 

\begin{corollary}\label{corollary:FL-Closure-Rules}
For every formula $\varphi$, the set of formulas $\FL(\varphi)$ contains $\varphi$, is closed under subformulas and has the following additional closure properties.
\begin{enumerate}
\item $\phantom{a}\diam{\pi_1 \cup \pi_2}^m \psi \in \FL(\varphi) \phantom{a} \Rightarrow \phantom{a} \diam{\pi_1}^m \psi \in \FL(\varphi) \And \diam{\pi_2}^m \psi \in \FL(\varphi)$,
\item $\phantom{a}\diam{\pi_1 ; \pi_2}^m \psi \in \FL(\varphi) \phantom{a} \Rightarrow \phantom{a} \diam{\pi_1}^{m-i} \diam{\pi_2}^{i}\psi \in \FL(\varphi)\phantom{a}$ for all $i \in \{ 0, \dots, m \}$,
\item $\phantom{a}\diam{\pi^\ast}^m \psi \in \FL(\varphi) \phantom{a}\Rightarrow\phantom{a} \diam{\pi}^{n} \diam{\pi^\ast}^{n'}\psi \in \FL(\varphi)\phantom{a}$ for all $n + n' \leq m$,
\item $\phantom{a}\diam{\rho?}^m\psi \in \FL(\varphi) \phantom{a}\Rightarrow\phantom{a} \rho \in \FL(\varphi)$.
\end{enumerate}
\end{corollary}
\begin{proof}
It is straightforward by Definition~\ref{definition:FL-closure} to show (by mutual induction) that $\varphi' \in \FL(\varphi)$ for all subformulas $\varphi'$ of $\varphi$ and $\diam{\pi}^m \varphi \in \FL^\Diamond\big(\diam{\pi}^m \varphi\big)$ always hold. With this, closure under subformulas easily follows from Lemma~\ref{lemma:FL-Closure-Rules}(i): If $\psi \in \FL(\varphi)$ and $\psi'$ is a subformula of $\psi$, then $\psi' \in \FL(\psi) \subseteq \FL(\varphi)$. Conditions \emph{(1)}--\emph{(4)} also directly follow from the above fact together with Lemma~\ref{lemma:FL-Closure-Rules}(i), we prove \emph{(3)} for the sake of illustration. If $\diam{\pi^\ast}^m \psi \in \FL(\varphi)$ then by Lemma~\ref{lemma:FL-Closure-Rules}(i) we find $\FL^\Diamond\big(\diam{\pi^\ast}^m \psi\big) \subseteq \FL(\varphi)$. Therefore, we have $\diam{\pi}^n \diam{\pi^\ast}^{n'} \psi \in \FL^\Diamond\big(\diam{\pi}^n \diam{\pi^\ast}^{n'} \psi\big) \subseteq \FL^\Diamond\big(\diam{\pi^\ast}^m \psi\big) \subseteq \FL(\varphi)$ as desired.    
\end{proof}

In the next subsection, we show that $\NPDL$ has the bounded finite model property, by defining appropriate \emph{filtrations} of models through the Fischer-Ladner closure.

\subsection{Filtration and the Bounded Finite Model Property}\label{subsection:FiltrationFMP}
In order to prove the bounded FMP (Corollary~\ref{corollary:StrongFMP}), we use \emph{filtrations} similarly to standard $\PDL$ (\emph{e.g.}, see \cite[Subsection 6.2]{HarelKozen2000}). However, to accommodate for costs we need to define the appropriate \emph{weighted} accessibility relations (also see many-valued $\PDL$ \cite[Definition 4.3]{Sedlar2020}). 

\begin{definition}[Filtrations]\label{definition:Filtration}
Let $\mathfrak{M} = \langle X, \{R_a \}_{a \in \AtProg}, \Val \rangle$ be a model and $\varphi$ be a formula. The \emph{filtration of $\mathfrak{M}$ through $\FL(\varphi)$} is the model
$
\mathfrak{M}_\varphi^f = \langle X^f, \{R_a^f\}, \Val^f \rangle
$ defined as follows. \\
The set $X^f := X/_{\leftrightsquigarrow_{\FL(\varphi)}}$ is the quotient of $X$ by the equivalence relation
$$
x \leftrightsquigarrow_{\FL(\varphi)} x'  \quad \text{ if and only if } \quad \forall \psi \in \FL(\varphi) \colon \mathfrak{M}, x \Vdash \psi \phantom{i}\Leftrightarrow\phantom{i} \mathfrak{M}, x' \Vdash \psi.
$$
The weighted relations $R_a^f \colon (X^f)^2 \to \mathbb{N}^\infty$ for $a \in \AtProg$ are defined via 
$$
R_a^f\big([x], [y]\big) = \bigwedge \{ R_a(x',y') \mid x' \in [x] \And y'\in [y] \}.
$$
The valuation is given by $\Val^f(p) = \{ [x] \mid \mathfrak{M}, x \Vdash p \}$ for $p \in \FL(\varphi)$, otherwise $\Val^f(q) = \varnothing$.
\end{definition}

We proceed with the central result of this subsection. The key `ingredients' in the proof are the $\FL$-closure properties of Corollary~\ref{corollary:FL-Closure-Rules} together with the validities of Proposition~\ref{proposition:ReductionAxiomsValid}.

\begin{theorem}[Filtration Theorem]\label{theorem:FiltrationTheorem}
Let $\mathfrak{M}_\varphi^f$ be the filtration of the model $\mathfrak{M}$ as in Definition~\ref{definition:Filtration}.
\begin{enumerate}[label = (\roman*)] 
\item For all formulas $\psi$ with $\psi \in \FL(\varphi)$ it holds that
$
\mathfrak{M}, x \Vdash \psi \ \Leftrightarrow \ \mathfrak{M}_\varphi^f, [x] \Vdash \psi.
$
\item For all programs $\pi$ with $\diam{\pi}^m\psi \in \FL(\varphi)$ the following hold:
\begin{enumerate}[label = (\alph*)]
\item $R_\pi^f\big([x],[y]\big) \leq R_\pi(x,y)$,
\item If $R_\pi^f\big([x],[y]\big) \leq m$ and $\mathfrak{M}, y \Vdash \psi$, then $\mathfrak{M}, x \Vdash \diam{\pi}^m\psi$. 
\end{enumerate}
\end{enumerate}
\end{theorem}

\begin{proof}
Simultaneous induction on the formula $\psi$ and program $\pi$ (we abbreviate $\mathfrak{M}^f:=\mathfrak{M}_\varphi^f$). 

\textbf{Formula induction.} The base case $p \in \Prop$ is covered by the definition of $\Val^f$ and the cases $\neg \psi$ and $\psi_1 \wedge \psi_2$ are standard induction arguments (note $\neg\psi \in \FL(\varphi) \Rightarrow \psi \in \FL(\varphi)$ and $\psi_1 \wedge \psi_2 \in \FL(\varphi) \Rightarrow \psi_1, \psi_2 \in \FL(\varphi)$ due to Corollary~\ref{corollary:FL-Closure-Rules}).

\emph{$\bullet$ Case $\diam{\pi}^m \psi$} :  
If $\mathfrak{M}, x \Vdash \diam{\pi}^m\psi$, then there is $y \in X$ with $R_\pi(x,y) \leq m$ and $\mathfrak{M}, y \Vdash \psi$. The inductive hypothesis \emph{(a)} on $\pi$ (note that we assume $\diam{\pi}^m\psi \in \FL(\varphi)$) gives $R^f_\pi\big([x], [y]\big) \leq R_\pi(x,y) \leq m$ and the inductive hypothesis \emph{(i)} on $\psi$ (note that $\psi \in \FL(\varphi)$ by Corollary~\ref{corollary:FL-Closure-Rules}) gives $\mathfrak{M}^f, [y] \Vdash \psi$. We thus showed $\mathfrak{M}^f, [x] \Vdash \diam{\pi}^m\psi$ as desired. 

Conversely, if $\mathfrak{M}^f, [x] \Vdash \diam{\pi}^m \psi$, then there is $[y] \in X^f$ such that $R_\pi^f\big([x], [y]\big) \leq m$ and $\mathfrak{M}^f, [y] \Vdash \psi$. From the inductive hypothesis \emph{(i)} on $\psi$ we get $\mathfrak{M}, y \Vdash \psi$ and subsequently the inductive hypothesis \emph{(b)} on $\pi$ yields $\mathfrak{M},x \Vdash \diam{\pi}^m \psi$ as desired. 

\textbf{Program induction.} 
\emph{$\bullet$ Case $a \in \AtProg$} : Let $\diam{a}^m\psi \in \FL(\varphi)$. Condition \emph{(a)} holds by Definition~\ref{definition:Filtration} since 
$R^f_a([x],[y]) = \bigwedge\{ R_a(x', y') \mid x' \in [x] \And y' \in [y] \} \leq R_a(x,y)$.
\\
For \emph{(b)}, suppose $R^f_a\big([x],[y]\big) \leq m$ and $\mathfrak{M}, y \Vdash \psi$. The former means that there are $x' \in [x], y' \in [y]$ such that $R_a(x',y') \leq m$ and the latter yields $\mathfrak{M}, y' \Vdash \psi$. Therefore, $\mathfrak{M}, x' \Vdash \diam{a}^m \psi$ and since we assume $\diam{a}^m\psi \in \FL(\varphi)$ we lastly obtain $\mathfrak{M}, x \Vdash \diam{a}^m \psi$ as needed.

\emph{$\bullet$ Case $\pi_1 \cup \pi_2$} : Let $\diam{\pi_1 \cup \pi_2}^m\psi \in \FL(\varphi)$. Condition \emph{(a)} holds because 
\begin{align*}
R_{\pi_1 \cup \pi_2}^f\big([x],[y]\big) & \ = \ R^f_{\pi_1}\big([x],[y]\big) \wedge R^f_{\pi_2}\big([x],[y]\big) \\  & \ \leq R_{\pi_1}(x,y) \wedge R_{\pi_2}(x,y) \ = \ R_{\pi_1 \cup \pi_2}(x,y) 
\end{align*}
by inductive hypothesis \emph{(a)} on $\pi_1$ and $\pi_2$ (note that by Corollary~\ref{corollary:FL-Closure-Rules}(1) we have $\diam{\pi_i}^m \psi \in \FL(\varphi)$).
\\
For \emph{(b)}, suppose $R^f_{\pi_1 \cup \pi_2}\big([x],[y]\big) \leq m$ and $\mathfrak{M}, y \Vdash \psi$. If $R^f_{\pi_1}\big([x],[y]\big) \leq m$ then the inductive hypothesis \emph{(b)} on $\pi_1$ yields $\mathfrak{M}, x \Vdash \diam{\pi_1}^m\psi$ and thus also $\mathfrak{M}, x \Vdash \diam{\pi_1 \cup \pi_2}^m\psi$ by Proposition~\ref{proposition:ReductionAxiomsValid}(4). If instead $R^f_{\pi_2}\big([x],[y]\big) \leq m$, the argument is completely analogous. 

\emph{$\bullet$ Case $\pi_1; \pi_2$} : Let $\diam{\pi_1;\pi_2}^m\psi \in \FL(\varphi)$. For \emph{(a)} we simply observe that 
$$
\bigwedge_{[v] \in X^f} R^f_{\pi_1}\big([x],[v]\big) + R^f_{\pi_2}\big([v],[y]\big) \leq \bigwedge_{v \in X} R_{\pi_1}(x,v) + R_{\pi_2}(v,y) 
$$     
due to the inductive hypothesis \emph{(a)} on $\pi_1$ and $\pi_2$ (note that by Corollary~\ref{corollary:FL-Closure-Rules}(2) we have $\diam{\pi_1}^m\diam{\pi_2}^0\psi \in \FL(\varphi)$ and by subformula closure $\diam{\pi_2}^0 \psi \in \FL(\varphi)$).
\\
For \emph{(b)}, suppose $R_{\pi_1;\pi_2}^f\big([x],[y]\big) \leq m$ and $\mathfrak{M}, y \Vdash \psi$. Due to the former, there exists $[v] \in X^f$ such that $R^f_{\pi_1}\big([x],[v]\big) + R^f_{\pi_2}\big([v],[y]\big) \leq m$. Let $i := R_{\pi_2}^f\big([v],[y]\big)$. Since $R_{\pi_2}^f\big([v],[y]\big) \leq i$ and $\mathfrak{M}, y \Vdash \psi$, we can use the inductive hypothesis \emph{(b)} on $\pi_2$ (note that by Corollary~\ref{corollary:FL-Closure-Rules}(2) we have $\diam{\pi_1}^{m-i} \diam{\pi_2}^{i} \psi \in \FL(\varphi)$ and $\diam{\pi_2}^{i}\psi \in \FL(\varphi)$) to obtain $\mathfrak{M}, v \Vdash \diam{\pi_2}^{i}\psi$. With $R_{\pi_1}^f\big([x], [v]\big) \leq m-i$ and the inductive hypothesis \emph{(b)} on $\pi_1$, this yields $\mathfrak{M}, x \Vdash \diam{\pi_1}^{m-i} \diam{\pi_2}^{i}\psi$ and therefore $\mathfrak{M}, x \Vdash \diam{\pi_1;\pi_2}^{m}\psi$ by Proposition~\ref{proposition:ReductionAxiomsValid}(5).     

\emph{$\bullet$ Case $\pi^\ast$} : Let $\diam{\pi^\ast}^m\psi \in \FL(\varphi)$. For \emph{(a)} it suffices to show $(R^f_\pi)^{[k]}\big([x],[y]\big) \leq (R_
{\pi})^{[k]}(x,y)$ for all $k \geq 0$, which is done by induction. The case $k = 0$ is straightforward and for the inductive step $k \mapsto k+1$ similarly to the case of composition above, one uses
$$
\bigwedge_{[v] \in X^f} R^f_{\pi}\big([x],[v]\big) + (R^f_{\pi})^{[k]}\big([v],[y]\big) \leq \bigwedge_{v \in X} R_{\pi}(x,v) + (R_{\pi})^{[k]}(v,y)
$$  
which follows from the inductive hypothesis on $k$ and inductive hypothesis \emph{(a)} on $\pi$ (note that by Corollary~\ref{corollary:FL-Closure-Rules}(3) we have $\diam{\pi}^m \diam{\pi^\ast}^0 \in \FL(\varphi)$).
\\
For \emph{(b)}, suppose $R^f_{\pi^\ast}\big([x], [y]\big) \leq m$ and $\mathfrak{M},y \Vdash \psi$. By the former, $(R^f_\pi)^{[k]} \leq m$ for some $k \geq 0$. If $(R_\pi^f)^{[0]}\big([x] , [y]\big) \leq m$ then $[x] = [y]$ implies $\mathfrak{M}, x \Vdash \psi$ implies $\mathfrak{M}, x \Vdash \diam{\pi^\ast}^m \psi$. If $(R^f_\pi)^{[1]}\big([x],[y]\big) \leq m$ note that $\mathfrak{M}, y \Vdash \psi$ implies $\mathfrak{M}, y \Vdash \diam{\pi^\ast}^0 \psi$ and we use the inductive hypothesis \emph{(b)} on $\pi$ (note that by Corollary~\ref{corollary:FL-Closure-Rules}(3) we have $\diam{\pi}^m \diam{\pi^\ast}^0 \psi \in \FL(\varphi)$) to obtain $\mathfrak{M}, x \Vdash \diam{\pi}^m \diam{\pi^\ast}^0 \psi$ which implies $\mathfrak{M}, x \Vdash \diam{\pi^\ast}^m \psi$ by Proposition~\ref{proposition:ReductionAxiomsValid}(6). If $k \geq 2$ then there are states $v_1, \dots, v_{k-1} \in X$ such that 
$$
m_1 + m_2 + \dots + m_k \phantom{i}:= \phantom{i} R^f_{\pi}\big([x], [v_1]\big) + R^f_{\pi}\big([v_1], [v_2]\big) + \dots + R^f_{\pi}\big([v_k], [y]\big) \phantom{i}\leq\phantom{i} m.    
$$      
We again have $\mathfrak{M}, y \Vdash \diam{\pi^\ast}^0 \psi$ and since $R^f_{\pi}\big([v_{k-1}], [y]\big) \leq m_k$ together with the inductive hypothesis \emph{(b)} on $\pi$ (note that by Corollary~\ref{corollary:FL-Closure-Rules}(3) we have $\diam{\pi}^{m_k} \diam{\pi^\ast}^0 \psi \in \FL(\varphi)$) we get $\mathfrak{M}, v_{k-1} \Vdash \diam{\pi}^{m_k} \diam{\pi^\ast}^0 \psi$, which implies $\mathfrak{M}, v_{k-1} \Vdash \diam{\pi^\ast}^{m_k}\psi$. Using $R_\pi^f\big([v_{k-2}], [v_{k-1}]\big) \leq m_{k-1}$ and the inductive hypothesis \emph{(b)} on $\pi$ (note that $\diam{\pi}^{m_{k-1}} \diam{\pi^\ast}^{m_k} \psi \in \FL(\varphi)$) we get $\mathfrak{M}, v_{k-2} \Vdash \diam{\pi}^{m_{k-1}} \diam{\pi^\ast}^{m_k} \psi$, which implies that $\mathfrak{M}, v_{k-2} \Vdash \diam{\pi^\ast}^{m_{k-1} + m_k} \psi$. Proceeding like this (note $\diam{\pi}^{m_j} \diam{\pi^\ast}^{m_{j+1} + \dots + m_k} \psi \in \FL(\varphi)$ for all $1 \leq j \leq k$), we eventually arrive at $\mathfrak{M}, x \Vdash \diam{\pi^\ast}^{m_1 + \dots + m_k} \psi$, which by Proposition~\ref{proposition:ReductionAxiomsValid}(3) implies $\mathfrak{M}, x \Vdash \diam{\pi^\ast}^m \psi$ as desired.  

\emph{$\bullet$ Case $\rho?$}: To see \emph{(a)} recall that $R_{\rho?}(x,y) = 0$ if and only if $x = y$ and $\mathfrak{M}, x \Vdash \rho$ (otherwise \emph{(a)} is trivial as $R_{\rho?}(x,y) = \infty$). By the inductive hypothesis \emph{(i)} on $\rho$ (note that by Corollary~\ref{corollary:FL-Closure-Rules}(4) we have $\rho \in \FL(\varphi)$) this implies $\mathfrak{M}^f, [x] \Vdash \rho$ and thus $R_{\rho?}^f\big([x],[y]\big) = 0$.
\\
For \emph{(b)}, let $R_{\rho?}^f\big([x],[y]\big) \leq m$ and $\mathfrak{M}, y \Vdash \psi$. The former yields $R_{\rho?}^f\big([x],[y]\big) = 0$, which holds if and only if $[x] = [y]$ and $\mathfrak{M}^f, [y] \Vdash \rho$. By inductive hypothesis \emph{(i)} on $\rho$ this implies $\mathfrak{M}, y \Vdash \rho$. Thus $\mathfrak{M}, y \Vdash \rho \wedge \psi$, which is equivalent to $\mathfrak{M}, y \Vdash \diam{\rho?}^m\psi$ by Proposition~\ref{proposition:ReductionAxiomsValid}(7).    
\end{proof} 

The bounded finite model property now easily follows.

\begin{corollary}[Bounded FMP]\label{corollary:StrongFMP}
The $\NPDL$-formula $\varphi$ is satisfiable if and only if it is satisfiable in some finite model of size at most $2^{\norm{\varphi}}$.
\end{corollary}
\begin{proof}
If $\varphi$ is satisfied in a model $\mathfrak{M}$, then as a consequence of Theorem~\ref{theorem:FiltrationTheorem} it is also satisfied in the filtration $\mathfrak{M}_\varphi^f$. By Lemma~\ref{lemma:FL-Finite} we know $|\FL(\varphi)| \leq \norm{\varphi}$ and it is easy to see that $|X^f| \leq 2^{|\FL(\varphi)|}$ by Definition~\ref{definition:Filtration} (since $[x] \mapsto \{ \psi \in \FL(\varphi) \mid \mathfrak{M}, x \Vdash \psi \}$ is injective).    
\end{proof}

However, unlike in standard $\PDL$, this does \emph{not} immediately yield decidability. Indeed, since we interpret formulas on $\mathbb{N}^\infty$-weighted graphs, there are still infinitely many `candidates' for weights of atomic programs $R_a$, even on a finite set. Therefore, we need to show that there also exists a \emph{bound on weights} sufficient to check the satisfiability of a formula.

\subsection{Bounding Weights}\label{subsection:TowardsDecidability}

In this subsection, for every formula $\varphi$ we aim to find a natural number $N(\varphi)$ such that satisfiability of $\varphi$ can be checked by only considering models with $R_a \colon X^2 \to \{ 0, 1, \dots, N(\varphi)+1\}$ for all atomic programs. Note that together with the bounded finite model property proved in the previous subsection (Corollary~\ref{corollary:StrongFMP}), this will establish decidability.

Intuitively, the bound $N(\varphi)$ is the largest $m$ for which some expression of the form `$\diam{\pi}^m$' appears (anywhere) in the formula $\varphi$. We make this precise in the following.  

\begin{definition}\label{definition:BoundOnWeights}
By mutual induction on the formula $\varphi \in \Formu$ and the program $\pi \in \Prog$, we define the natural numbers $N(\varphi)$ and  $N(\pi)$ as follows.\\
\textbf{\textit{Formulas}:}
\begin{itemize}
\item $N(p) = 0$ for $p \in \Prop$, 
\item $N(\neg\varphi) = N(\varphi)$,
\item $N(\varphi_1 \wedge \varphi_2) = \mathrm{max}\{N(\varphi_1), N(\varphi_2)\}$,
\item $N(\diam{\pi}^m \varphi) = \mathrm{max}\{m, N(\varphi), N(\pi) \}$.
\end{itemize}
\textbf{\textit{Programs}:}
\begin{itemize}
\item $N(a) = 0$ for $a \in \AtProg$,
\item $N(\pi_1 \cup \pi_2) = \mathrm{max}\{ N(\pi_1), N(\pi_2) \}$,
\item $N(\pi_1 ; \pi_2) = \mathrm{max}\{ N(\pi_1), N(\pi_2) \}$, 
\item $N(\pi^\ast) = N(\pi)$, 
\item $N(\varphi?) = N(\varphi)$. 
\end{itemize}
\noindent (Since all maxima range over finite sets, these are well-defined natural numbers.)  
\end{definition}

The idea now is to turn a model into a model where only finitely many weights appear by `cutting off' all weights which go beyond a certain `threshold'.  

\begin{definition}[Cut-off models]\label{definition:BoundedWeightsModel}
Let $\mathfrak{M} = \langle X, \{ R_a \}_{a \in \AtProg}, \Val \rangle$ be a model and $n \in \mathbb{N}$. The corresponding \emph{$n$-cut-off model} $\mathfrak{M}^n = \langle X, \{ R_a^n \}_{a \in \AtProg}, \Val \rangle$ is defined by $R^n_a(x,y) = R_a(x,y) \wedge n$. 
\end{definition}

The main technical result of this subsection shows that satisfiability of $\varphi$ stays preserved as long as the cut-off happens (strictly) above $N(\varphi)$.

\begin{proposition}\label{proposition:BoundOnWeights}
Let $\mathfrak{M}$ be a model and $\{ \mathfrak{M}^n \}_{n \in \mathbb{N}}$ be the cut-off models as in Definition~\ref{definition:BoundedWeightsModel}.
\begin{enumerate}[label = (\roman*)]
\item For all formulas $\varphi \in \Formu$:
$
(\mathfrak{M}, x \Vdash \varphi \ \Leftrightarrow \ \mathfrak{M}^n , x \Vdash \varphi) \phantom{i}\text{ for all }n > N(\varphi).
$
\item For all programs $\pi \in \Prog$:
$ 
\phantom{a}
(R_\pi(x,y) \wedge n \phantom{i}=\phantom{i} R^n_\pi(x,y) \wedge n) \phantom{i}\text{ for all } n > N(\pi). 
$
\end{enumerate}
\end{proposition} 
\begin{proof}
Simultaneous induction on the formula $\varphi$ and the program $\pi$.

\textbf{Formula induction.} The base case $p \in \Prop$ is immediate from $\Val^n = \Val$ and the cases $\neg\varphi$ and $\varphi_1 \wedge \varphi_2$ are easily obtained from the corresponding inductive hypotheses \emph{(i)} on $\varphi$ and $\varphi_1, \varphi_2$, respectively (recall $N(\neg\varphi) = N(\varphi)$ and $N(\varphi_1 \wedge \varphi_2) \geq N(\varphi_i)$).  

\emph{$\bullet$ Case $\diam{\pi}^m \varphi$} :
Let $n > N\big(\diam{\pi}^m \varphi\big)$. Then $\mathfrak{M}, x \Vdash \diam{\pi}^m \varphi$ if and only if there is $y \in X$ with $\mathfrak{M}, y \Vdash \varphi$ and $R_\pi(x,y) \leq m$. By inductive hypothesis \emph{(i)} on $\varphi$ (note $n > N(\varphi)$) we have 
$$
\mathfrak{M}, y \Vdash \varphi \phantom{i}\text{ if and only if }\phantom{i} \mathfrak{M}^n, y \Vdash \varphi
$$ 
and by the inductive hypothesis \emph{(ii)} on $\pi$ (note $n > N(\pi)$) and the fact that $n > m$ we have 
$$
R_\pi(x,y) \leq m \phantom{i}\text{ if and only if }\phantom{i} R^n_\pi(x,y) \leq m.
$$ 
Thus $\mathfrak{M}, x \Vdash \diam{\pi}^m \varphi$ if and only if there exists some $y \in X$ with $\mathfrak{M}^n, y \Vdash \varphi$ and $R_\pi^n(x,y) \leq m$, which by definition is equivalent to $\mathfrak{M}^n, x \Vdash \diam{\pi}^m \varphi$.  

\textbf{Program induction.} For the base case $a \in \AtProg$, by Definition~\ref{definition:BoundedWeightsModel} we have $R^n_a(x,y) = R_a(x,y) \wedge n$ and thus $R^n_a(x,y) \wedge n = R_a(x,y) \wedge n$. 

\emph{$\bullet$ Case $\pi_1 \cup \pi_2$} : Let $n > N(\pi_1 \cup \pi_2) = \mathrm{max}\{ N(\pi_1), N(\pi_2) \}$. We compute 
\begin{align*}
R_{\pi_1 \cup \pi_2}(x,y) \wedge n \phantom{a}&= \phantom{a} R_{\pi_1}(x,y) \wedge R_{\pi_2}(x,y) \wedge n \\
& = \phantom{a} \big(R_{\pi_1}(x,y) \wedge n\big) \wedge \big(R_{\pi_2}(x,y) \wedge n\big) \\
& = \phantom{a} \big(R^n_{\pi_1}(x,y) \wedge n\big) \wedge \big(R^n_{\pi_2}(x,y) \wedge n\big) \\
&= \phantom{a} R^n_{\pi_1}(x,y) \wedge R^n_{\pi_2}(x,y) \wedge n \phantom{a}=\phantom{a} R_{\pi_1 \cup \pi_2}(x,y) \wedge n,
\end{align*}
where the third equality uses the inductive hypothesis \emph{(ii)} on $\pi_1$ and $\pi_2$ (note $n > N(\pi_i)$).

\emph{$\bullet$ Case $\pi_1 ; \pi_2$} : Let $n > N(\pi_1 ; \pi_2) = \mathrm{max}\{ N(\pi_1), N(\pi_2) \}$. We compute 
{\small
\begin{align*}
R_{\pi_1;\pi_2}(x,y) \wedge n \phantom{a} & = \phantom{i} \big(\bigwedge_{v \in X} R_{\pi_1}(x,v) + R_{\pi_2}(v,y)\big) \wedge n \\ 
& = \phantom{i} \bigwedge_{v \in X} \big((R_{\pi_1}(x,v) \wedge n) + (R_{\pi_2}(v,y) \wedge n)\big) \wedge n \\ 
& = \phantom{i} \bigwedge_{v \in X} \big((R^n_{\pi_1}(x,v) \wedge n) + (R^n_{\pi_2}(v,y) \wedge n)\big) \wedge n \\
& = \phantom{i} \big(\bigwedge_{v \in X} R^n_{\pi_1}(x,v) + R^n_{\pi_2}(v,y)\big) \wedge n \phantom{a}=\phantom{a} R_{\pi_1;\pi_2}^n(x,y) \wedge n, 
\end{align*}
}
the third `$=$' uses inductive hypothesis \emph{(ii)} on $\pi_1, \pi_2$ (note $n > N({\pi_i})$), the second and fourth ones use the laws $(\bigwedge b_i) \wedge n = \bigwedge (b_i \wedge n)$ and $(c+d) \wedge n = \big((c \wedge n) + (d \wedge n)\big) \wedge n$ in $\mathbb{N}^\infty$.     

\emph{$\bullet$ Case $\pi^\ast$} : Let $n > N(\pi^\ast) = N(\pi)$. It suffices to show that $(R_\pi)^{[k]}(x,y) \wedge n = (R^n_\pi)^{[k]} \wedge n$ holds for all $k \geq 0$, which is done by induction. The base case $k = 0$ is obvious since $(R_\pi)^{[0]} = (R^n_\pi)^{[0]}$ and the inductive step $k \mapsto k+1$ is similar to the case of sequential composition $;$ above, using the inductive hypotheses on $k$ and on $\pi$ (note $n> N(\pi)$).

\emph{$\bullet$ Case $\psi?$} : Let $n > N(\psi?) = N(\psi)$. By definitions we have
$$
R_{\psi?}(x,y) = \begin{cases}
0 & \text{if }  x = y \text{ \& } \mathfrak{M},x \Vdash \psi, \\ 
\infty & \text{otherwise.}
\end{cases}
$$
and
$$
R^n_{\psi?}(x,y) = \begin{cases}
0 & \text{if }  x = y \text{ \& } \mathfrak{M}^n,x \Vdash \psi, \\ 
\infty & \text{otherwise.}
\end{cases}
$$ 
By the inductive hypothesis \emph{(i)} on $\psi$ (note $n > N(\psi)$) we have $\mathfrak{M},x \Vdash \psi$ if and only if $\mathfrak{M}^n,x \Vdash \psi$, showing $R_{\psi?} = R^n_{\psi?}$ which trivially implies $R_{\psi?} \wedge n = R^n_{\psi?} \wedge n$. 
\end{proof}

Combining the bounded finite model property (Corollary~\ref{corollary:StrongFMP}) and the previous result about the bound on weights yields the following. 

\begin{theorem}[Decidability]\label{theorem:Decidability}
Satisfiability and validity of $\NPDL$ are decidable.   
\end{theorem}
\begin{proof}
Let $\varphi$ be a formula. By the bounded finite model property (Corollary~\ref{corollary:StrongFMP}) $\varphi$ and Proposition~\ref{proposition:BoundOnWeights}, $\varphi$ is satisfiable if and only if it is satisfiable in a finite model of size at most $2^{\norm{\varphi}}$ where the weights of atomic programs only take values in $\{0 , 1, \dots, N(\varphi) + 1 \}$. Lastly, note that only the finitely many propositional variables and atomic programs (plus finitely many complex programs computed from these) occurring in $\varphi$ are `relevant' to check  satisfiability. The fact that validity of $\varphi$ is decidable follows as usual from the fact that it is equivalent to checking satisfiability of $\neg\varphi$.    
\end{proof}

\begin{remark}[Adding converse]\label{remark:Converse}
The above decidability proof can be easily extended to $\NPDL$\emph{ with converse programs} $\pi^\smallsmile$ interpreted as $R_{\pi^\smallsmile}(x,y) = R_{\pi}(y,x)$, as it is straightforward to verify the validities $\varphi \rightarrow \boxm{\pi}^m \diam{\pi^\smallsmile}^m \varphi$ and $\varphi \rightarrow \boxm{\pi^\smallsmile}^m \diam{\pi}^m \varphi$ and program equivalences $(\pi_1 \cup \pi_2)^\smallsmile\cong \pi_1^\smallsmile \cup \pi_2^\smallsmile$, $(\pi_1;\pi_2)^\smallsmile \cong \pi_2^\smallsmile ; \pi_1^\smallsmile$, and $(\psi?)^\smallsmile \cong \psi?$ analogous to standard $\PDL$. 
\end{remark}

\section{Completeness}\label{section:Completeness}

In this section, we use the filtration technique developed in the previous section to obtain (weak) completeness via \emph{finite canonical models}. The axioms of $\NPDL$ are as follows. 

\begin{definition}[$\vdash_{\NPDL}$]\label{definition:NPDLProvability}
A formula $\varphi$ is \emph{provable}, denoted by $\vdash_{\NPDL} \varphi$, if it can be obtained from the following axioms together with propositional reasoning, modus ponens and the monotonicity rule $\varphi \rightarrow \psi \blacktriangleright \diam{\pi}^m \varphi \rightarrow \diam{\pi}^m \psi$ for every $m \in \mathbb{N}$:  
\begin{enumerate}
\item $\diam{\pi}^m \bot \phantom{i}\leftrightarrow\phantom{i} \bot$
\item $\diam{\pi}^m (\varphi \vee \psi) \phantom{i}\leftrightarrow\phantom{i} \diam{\pi}^m \varphi \vee \diam{\pi}^m \psi$
\item $\diam{\pi}^n \varphi \phantom{i}\rightarrow\phantom{i} \diam{\pi}^m \varphi \phantom{i}$ for all $n \leq m$
\item $\diam{\pi_1 \cup \pi_1}^m \varphi \phantom{i}\leftrightarrow\phantom{i} \diam{\pi_1}^m \varphi \vee \diam{\pi_2}^m \varphi$
\item $\diam{\pi_1 ; \pi_2}^m \varphi \phantom{i}\leftrightarrow\phantom{i} \bigvee_{i = 0}^m \diam{\pi_1}^{m-i} \diam{\pi_2}^i \varphi$
\item $\diam{\pi^\ast}^m \varphi \phantom{i}\leftrightarrow\phantom{i} \varphi \vee \bigvee_{i = 0}^{m} \diam{\pi}^{m-i} \diam{\pi^\ast}^{i}\varphi $
\item $\diam{\psi?}^m \varphi \phantom{i}\leftrightarrow\phantom{i} \psi \wedge \varphi$ 
\item $\big(\varphi \wedge \bigwedge_{i = 0}^m \boxm{\pi^\ast}^i(\varphi \rightarrow \boxm{\pi}^{m-i}\varphi) \big) \rightarrow \boxm{\pi^\ast}^m \varphi$
\end{enumerate}
\end{definition} 

Soundness of (1)--(7) was shown in Proposition~\ref{proposition:ReductionAxiomsValid}, we proceed to verify soundness of the induction rules (8). Let $\mathfrak{M} = \langle X, \{R_a\}_{a \in \AtProg}, \Val \rangle$ be a model and $x \in X$ such that $\mathfrak{M}, x \Vdash \varphi \wedge \bigwedge_{i = 0}^m \boxm{\pi^\ast}^i(\varphi \rightarrow \boxm{\pi}^{m-i}\varphi)$. To show $\mathfrak{M}, x \Vdash \boxm{\pi^\ast}^m\varphi$ we inductively show 
$$
R^{[k]}_\pi(x,y) \leq m \Rightarrow \mathfrak{M}, y \Vdash \varphi
$$
for all $k$. The case $k = 0$ follows from $\mathfrak{M}, x \Vdash \varphi$. For the inductive step $k \mapsto k+1$ suppose $R^{[k+1]}_\pi(x,y) \leq m$. Then there exist $v \in X$ such that $R^{[k+1]}_\pi(x,y) = R^{[k]}_\pi(x,v) + R_\pi(v,y)$. Set $i := R^{[k]}_\pi(x,v)$. Now we have $\mathfrak{M}, v \Vdash \varphi$ by inductive hypothesis (note $i \leq m$) and $\mathfrak{M}, v \Vdash \varphi \rightarrow \boxm{\pi}^{m-i}\varphi$ by the initial assumption. Thus we have $\mathfrak{M}, v \Vdash \boxm{\pi}^{m-i}\varphi$ and since $R_\pi(v,y) \leq m-i$ we obtain $\mathfrak{M}, y \Vdash \varphi$ as desired. 

We now define the \emph{canonical pseudo-model} based on the maximally consistent sets (`pseudo' indicates that it is not a model in the sense of Definition~\ref{definition:Model}). As usual we say that a set of formulas $\Gamma$ is \emph{consistent} if there are no formulas $\varphi_1, \dots, \varphi_\ell \in \Gamma$ such that $\vdash_{\NPDL} \bigwedge_{i = 1}^\ell \varphi_i \rightarrow \bot$ and \emph{maximally consistent} if $\Gamma \subseteq \Gamma'$ implies $\Gamma = \Gamma'$ for every other consistent set $\Gamma'$. 

\begin{definition}[Canonical pseudo-model]\label{definition:PseudoCanonicalModel} 
We define the $\NPDL$-\emph{canonical pseudo-model} $\mathfrak{M}^c = \langle X^c, \{ R^c_\pi \}_{\pi \in \Pi}, \Val^c \rangle$ as follows (note: we define $R^c_\pi$ for \emph{all} programs).
\begin{itemize}
\item $X^c$ is the collection of all maximally $\NPDL$-consistent sets,
\item $R^c_\pi(\Gamma, \Gamma') = \bigwedge\{ m \in \mathbb{N} \mid \forall \varphi \colon \varphi \in \Gamma' \Rightarrow \diam{\pi}^m\varphi \in \Gamma \}$,
\item $\Val^c(p) = \{ \Gamma \in X^c \mid p \in \Gamma \}$
\end{itemize}
\end{definition}

Similarly to the non-dynamic case \cite[Lemma 4.2]{Ma2015}, first we observe that `\emph{truth is membership}' in the canonical pseudo-model.

\begin{lemma}[Truth Lemma]\label{lemma:TruthLemma}
For all formulas $\varphi$, it holds that $\mathfrak{M}^c, \Gamma \Vdash \varphi \Leftrightarrow \varphi \in \Gamma$.  
\end{lemma}
\begin{proof}
Induction on the formula, the case of propositional variables is immediate by definition and the cases for propositional connectives are straightforward, so we consider the remaining case $\diam{\pi}^m \varphi$ below. \\
If $\mathfrak{M}^c, \Gamma \Vdash \diam{\pi}^m \varphi$ then there exists $\Gamma' \in X^c$ with $\mathfrak{M}^c, \Gamma' \Vdash \varphi$ and $R_\pi^c(\Gamma, \Gamma') \leq m$. The former together with the inductive hypothesis yields $\varphi \in \Gamma'$, which together with the latter implies $\diam{\pi}^m\varphi$ as desired (note that here we use the axiom $\diam{\pi}^n \varphi \rightarrow \diam{\pi}^m \varphi$ with $n = R^c_\pi(\Gamma, \Gamma')$). \\
Conversely, assume that $\diam{\pi}^m\varphi \in \Gamma$, we want to construct a maximally consistent theory $\Gamma'$ with $\varphi \in \Gamma'$ and $R_\pi^c(\Gamma, \Gamma') \leq m$. For this, it suffices to show that the set 
$$
\{ \varphi \} \cup \{ \psi \mid \boxm{\pi}^m\psi \in \Gamma \} 
$$   
is consistent. Suppose towards contradiction that it is not consistent, then there are $\boxm{\pi}^m\psi_1, \dots, \boxm{\pi}^m\psi_\ell \in \Gamma$ such that $\vdash_{\NPDL} \psi_1 \wedge \dots \wedge \psi_\ell \rightarrow \neg\varphi$. The monotonicity rule yields 
$$
\vdash_{\NPDL} \boxm{\pi}^m \psi_1 \wedge \boxm{\pi}^m\psi_\ell \rightarrow \neg\diam{\pi}^m\varphi.
$$
By construction we have that all $\boxm{\pi}^m\psi_j \in \Gamma$, therefore by the above $\neg\diam{\pi}^m\varphi$, a contradiction to our assumption that $\diam{\pi}^m\varphi \in \Gamma$ and $\Gamma$ is consistent.     
\end{proof}

The following shows that the canonical pseudo-model is `well-behaved' with respect to all program constructs except iteration.

\begin{proposition}\label{proposition:CanModel-Constructs}
The following hold in the canonical pseudo-model:
\begin{enumerate}
\item $R^c_{\pi_1 \cup \pi_2} = R^c_{\pi_1} \wedge R^c_{\pi_2}$,
\item $R^c_{\pi_1 ; \pi_2} = R^c_{\pi_1} \circ R^c_{\pi_2}$,
\item $R^c_{\psi?} = \Delta_{ \sem{\psi} }$.
\end{enumerate}
\end{proposition} 
\begin{proof}
\emph{(1)}: $R^c_{\pi_1}(\Gamma, \Gamma') \wedge R^c_{\pi_2}(\Gamma, \Gamma') \geq  R^c_{\pi_1 \cup \pi_2}(\Gamma, \Gamma')$ is clear by definition (and the reduction axiom for $\cup$). For the converse, suppose towards contradiction that $R^c_{\pi_1}(\Gamma, \Gamma') \wedge R^c_{\pi_2}(\Gamma, \Gamma') > R^c_{\pi_1 \cup \pi_2}(\Gamma, \Gamma') =: m$. Then there are $\boxm{\pi_1}^m \varphi, \boxm{\pi_2}^m \psi \in \Gamma$ such that $\varphi, \psi \notin \Gamma'$. However, we have $\boxm{\pi_1 \cup \pi_2}^m (\varphi \vee \psi) \in \Gamma$ and thus $\varphi \vee \psi \in \Gamma'$, contradicting maximality of $\Gamma'$.  
\\
\emph{(2)}: We want to show that 
$
R^c_{\pi_1 ; \pi_2}(\Gamma, \Gamma') = \bigwedge_{\Delta \in X^c} R^c_{\pi_1}(\Gamma, \Delta) + R^c_{\pi_2}(\Delta, \Gamma')
$. 
\\
To see `$\leq$' suppose the right hand side is given by $m := R^c_{\pi_1}(\Gamma, \Delta) + R^c_{\pi_2}(\Delta, \Gamma')$. Let $i := R^c_{\pi_1}(\Gamma, \Delta)$. Now $R^c_{\pi_1 ; \pi_2}(\Gamma, \Gamma') \leq m$ holds since for all formulas $\varphi$ we have 
$$
\varphi \in \Gamma' \ \Rightarrow \ \diam{\pi_2}^{m-i}\varphi \in \Delta \ \Rightarrow \ \diam{\pi_1}^i \diam{\pi_2}^{m-i} \varphi \in \Gamma \ \Rightarrow \ \diam{\pi_1; \pi_2}^m \varphi \in \Gamma.
$$
For the opposite inequality `$\geq$', now let $R^c_{\pi_1 ; \pi_2}(\Gamma, \Gamma') = m$. Note that it suffices to show that there exists some $i \in \{ 0, \dots, m \}$ such that the following set of formulas is consistent: 
$$
\Delta^i = \{ \varphi \mid [\pi_1]^i \varphi \in \Gamma\} \cup \{ \diam{\pi_2}^{m-i}\psi \mid \psi \in \Gamma' \}.
$$
Suppose towards contradiction that all $\Delta^i$ are inconsistent. Then for each $i$ we there are $[\pi_1]^i\varphi^i_1, \dots, [\pi_1]^i\varphi^i_{k_i} \in \Gamma$ and $\psi^i_1, \dots, \psi^i_{\ell_i} \in \Gamma'$ such that 
$$
\vdash_{\NPDL} \varphi^i_1 \wedge \dots \wedge \varphi^i_{k_i} \wedge \diam{\pi_2}^{m-i}\psi^i_1 \wedge \dots \wedge \diam{\pi_2}^{m-i}\psi^i_{\ell_i} \rightarrow \bot.
$$
Set $\Phi^i = \bigwedge \{\varphi^i_j \mid 1 \leq j \leq k_i\}$ and $\Psi^i = \bigwedge\{ \psi^i_j \mid 1 \leq j \leq \ell_i \}$. With this and the fact that $\vdash_{\NPDL} \diam{\pi_2}^{m-i}(p \wedge q) \rightarrow \diam{\pi_2}^{m-i} p \wedge \diam{\pi_2}^{m-i} q$ the above implies 
$\vdash_{\NPDL} \Phi^i \rightarrow \neg \diam{\pi_2}^{m-i}\Psi^i$, thus by monotonicity of $\boxm{\pi_1}^i$ we find 
$$
\vdash_{\NPDL} \boxm{\pi_1}^i\Phi^i \rightarrow \neg \diam{\pi_1}^i\diam{\pi_2}^{m-i}\Psi^i.
$$ 
Now let $\Psi := \bigwedge \{ \Psi^i \mid 0 \leq i \leq m \}$. Since $\Psi \rightarrow \Psi^i$ the above also yields 
$$
\vdash_{\NPDL} \boxm{\pi_1}^i\Phi^i \rightarrow \neg \diam{\pi_1}^i\diam{\pi_2}^{m-i}\Psi.
$$ 
Now, on the one hand, observe that $\Psi \in \Gamma'$ and therefore it holds that $\diam{\pi_1 ; \pi_2}^m \Psi \in \Gamma$ (since $R^c_{\pi_1; \pi_2}(\Gamma, \Gamma') = m$). Due to  our axioms, we get $\bigvee_{i = 0}^m \diam{\pi_1}^i \diam{\pi_2}^{m-i} \Psi \in \Gamma$ and therefore $\diam{\pi_1}^i \diam{\pi_2}^{m-1}\Psi \in \Gamma$ must hold for some $i$. On the other hand, however, since by definition $\diam{\pi_1}^i \Phi^i \in \Gamma$, the above derivation implies $\neg\diam{\pi_1}^i \diam{\pi_2}^{m-i}\Psi \in \Gamma$, a contradiction. 
\\
\emph{(3)}: Note that $\diam{\psi?}^m \varphi \in \Gamma$ if and only if $\diam{\psi?}^0 \varphi \in \Gamma$ due to our reduction axioms for tests. Therefore either $R^c_{\psi?}(\Gamma, \Gamma') = \infty$ or $R^c_{\psi?}(\Gamma, \Gamma') = 0$. The latter holds if and only if $\forall \varphi \colon \varphi \in \Gamma' \Rightarrow \psi \wedge \varphi \in \Gamma$ which is equivalent to $\Gamma' = \Gamma$ (by maximality and $\Gamma' \subseteq \Gamma$) and $\mathfrak{M}^c, \Gamma \Vdash \psi$ (by the Lemma~\ref{lemma:TruthLemma} and $\psi \in \Gamma$). 
\end{proof}

While the canonical pseudo-model does \emph{not} satisfy $R^c_{\pi^\ast} = (R^c_{\pi})^\ast$, it still retains some similar properties, as shown in the following. 

\begin{lemma}\label{lemma:CanonicalModelAstProperties}
For every program $\pi$, the canonical pseudo-model satisfies:
\begin{enumerate}
    \item $R^c_{\pi^\ast}(\Gamma, \Gamma) = 0$,
    \item $R^c_\pi(\Gamma, \Gamma') \geq R^c_{\pi^\ast}(\Gamma, \Gamma')$, 
    \item $R^c_{\pi^\ast}(\Gamma, \Delta) + R^c_{\pi^\ast}(\Delta, \Gamma') \geq R^c_{\pi^\ast}(\Gamma, \Gamma')$.
\end{enumerate}
\end{lemma} 
\begin{proof}
\emph{(1)}: If $\varphi \in \Gamma$, then also $\diam{\pi^\ast}^0\varphi \in \Gamma$ since $\vdash_{\NPDL} \varphi \rightarrow \diam{\pi^\ast}^0\varphi$.

\emph{(2)}: Let $R^c_{\pi}(\Gamma, \Gamma') = m$. If $\varphi \in \Gamma'$, then $\diam{\pi}^m\varphi \in \Gamma$, since the monotonicity rule yields $\vdash_{\NPDL} \diam{\pi}^m \varphi \rightarrow \diam{\pi}^m \diam{\pi^\ast}^0\varphi$ we find $\diam{\pi}^m \diam{\pi^\ast}^0 \varphi \in \Gamma$, which lastly yields $\diam{\pi^\ast}^m \varphi \in \Gamma$, showing $R^c_{\pi^\ast}(\Gamma, \Gamma') \leq m$ as desired.

\emph{(3)}: Let $R^c_{\pi^\ast}(\Gamma, \Delta) = i$ and $R^c_{\pi^\ast}(\Delta, \Gamma') = m-i$. If $\varphi \in \Gamma'$ then $\diam{\pi^\ast}^{m-i}\varphi \in \Delta$ and therefore $\diam{\pi^\ast}^i\diam{\pi^\ast}^{m-i}\varphi \in \Gamma$. Due to $\vdash_{\NPDL} \diam{\pi^\ast}^i\diam{\pi^\ast}^{m-i}\varphi \rightarrow \diam{\pi^\ast}^{m}\varphi$ we get $\diam{\varphi}^m\varphi \in \Gamma$ as well, showing $R^c_{\pi^\ast}(\Gamma, \Gamma') \leq m$ as desired.    
\end{proof}

Crucially, the Filtration Theorem (Theorem~\ref{theorem:FiltrationTheorem}) still works for the canonical pseudo-model. 

\begin{theorem}[Filtration Theorem II]\label{theorem:FiltrationTheorem-II}
Let $\mathfrak{M}_\varphi^f$ be the filtration of the canonical pseudo-model $\mathfrak{M}^c$ as in Definition~\ref{definition:Filtration}.
\begin{enumerate}[label = (\roman*)] 
\item For all formulas $\psi$ with $\psi \in \FL(\varphi)$:
$
\phantom{i}\mathfrak{M}^c, x \Vdash \psi \ \Leftrightarrow \ \mathfrak{M}_\varphi^f, [x] \Vdash \psi.
$
\item For all programs $\pi$ with $\diam{\pi}^m\psi \in \FL(\varphi)$:
\begin{enumerate}[label = (\alph*)]
\item $R_\pi^f\big([x],[y]\big) \leq R^c_\pi(x,y)$,
\item If $R_\pi^f\big([x],[y]\big) \leq m$ and $\mathfrak{M}^c, y \Vdash \psi$, then $\mathfrak{M}^c, x \Vdash \diam{\pi}^m\psi$. 
\end{enumerate}
\end{enumerate}
\end{theorem}

\begin{proof}
The only part that is not completely analogous to the proof of Theorem~\ref{theorem:FiltrationTheorem} is to show that \emph{(a)} holds for iteration. We show by induction that for all $m\in \mathbb{N}$ 
$$
R^c_{\pi^\ast}(x,y) = m \quad \Rightarrow \quad R_{\pi^\ast}^f([x],[y]) \leq m. 
$$
We begin with the case $m = 0$. Let $R^c_{\pi^\ast}(x,y) = 0$. Define 
$$
E = \{ z \mid R_{\pi^\ast}^f([x],[z]) = 0 \}
$$  
and note that there exists a formula $\chi_E$ such that $\mathfrak{M}^c, z \Vdash \chi_E \Leftrightarrow z \in E$ since $FL(\varphi)$ is finite. We want to show $y \in E$, for which it suffices to show that $x \Vdash \boxm{\pi^\ast}^0 \chi_E$. This we derive from the induction axiom for $[\pi^\ast]^0$. We have $x \Vdash \chi_E$ since $R^f_{\pi^\ast}([x],[x]) = 0$. Now whenever $R^c_{\pi^\ast}(x,z) = 0$ and $z \Vdash \chi_E$ and $R^c_\pi(z,z') = 0$ we have $R^f_{\pi^\ast}([x],[z']) \leq R^f_{\pi^\ast}([x],[z]) + R^f_{\pi}([z],[z']) = 0$ (since $z \Vdash \chi_E$ and inductive hypothesis on $\pi$), thus $z' \in E$. So we showed $x \Vdash \chi_E \wedge \boxm{\pi^\ast}^0(\chi_E \rightarrow \boxm{\pi}^0\chi_E)$ as needed. 

Now inductively assume $R^c_{\pi^\ast}(u,v) = n \Rightarrow R_{\pi^\ast}^f([u],[v]) \leq n$ holds for all $n < m$ and let $R^c_{\pi^\ast}([x],[y]) = m$. Similarly to before, define
$
E = \{ z \mid R_{\pi^\ast}^f([x],[z]) \leq R^c_{\pi^\ast}(x,z) \}
$
and let $\chi_E$ be a formula defining this set. As before, we have $x \Vdash \chi_E$, so it suffices to show $x \Vdash \boxm{\pi^\ast}^i (\chi_E \rightarrow \boxm{\pi}^{m-i}\chi_E)$ for all $i = 0, \dots, m$, since then the induction axiom for $[\pi^\ast]^m$ yields $x \Vdash \boxm{\pi^\ast}^m \chi_E$ and thus $y \Vdash \chi_E$ as needed. Let $R^c_{\pi^\ast}(x,z) \leq i$ and $z \Vdash \chi_E$ and $R^c_\pi(z,z') \leq m-i$. Note that $R^c_{\pi^\ast}(x,z') \leq R^c_{\pi^\ast}(x,z) + R^c_\pi(z,z')  \leq m$. If $R^c_{\pi^\ast}(x,z') = n < m$, then the inductive hypothesis on $n$ yields $R_{\pi^\ast}^f([x],[z']) \leq n$, showing $z' \Vdash \chi_E$. If $R^c_{\pi^\ast}(x,z') = m$ then 
$$
R_{\pi^\ast}^f([x],[z']) \leq R^f_{\pi^\ast}([x],[z]) + R_\pi^f([z],[z']) \leq m 
$$ 
where we used $z \Vdash \chi_E$ and the inductive hypothesis on $\pi$ as $R_\pi^f([z],[z']) \leq R^c_{\pi}(z, z')$. This finishes the proof.
\end{proof}

Since $\mathfrak{M}^f$ is a (proper) model by definition, completeness is now a direct consequence.  

\begin{corollary}[Completeness]\label{corollary:completeness}
A formula is $\NPDL$-valid if and only if it is $\vdash_{\NPDL}$-provable.
\end{corollary}
\begin{proof}
Soundness was discussed before, completeness is shown as follows. If $\not\vdash_{\NPDL} \varphi$, then $\{\neg\varphi\}$ is consistent, thus (by Lindenbaum's Lemma) can be extended to a maximally consistent set $\Gamma$. By Lemma~\ref{lemma:TruthLemma} and Theorem~\ref{theorem:FiltrationTheorem-II} we get that $\mathfrak{M}^f_\varphi, [\Gamma] \not\Vdash \varphi$, showing that $\varphi$ is not valid.      
\end{proof}

\section{Complexity Bounds}\label{section:Translations}
In this section, we obtain bounds on the complexity $\NPDL$ using translation between this logic and standard $\PDL$. More specifically, we obtain a \textsc{exptime}-lower bound via a translation $\sigma$ of standard $\PDL$ into $\NPDL$ (Corollary~\ref{Corollary:ComplexityLowerBound}) and we obtain a \textsc{2exptime}-upper bound via a translation of $\NPDL$ into standard $\PDL$ over an extended set of atomic programs (Corollary~\ref{Corollary:ComplexityUpperBound}). We also give a complexity bound for the model checking problem of $\NPDL$ which significantly improves the bound obtained via the translation into $\PDL$ (Subsection~\ref{subsection:ModelCheckingComplexity}). 

\subsection{Satisfiability Problem via Logical Translations}\label{subsection:SatisfiabilityComplexityTranslations}

Translating standard $\PDL$ into $\NPDL$ is fairly straightforward, recalling that the weight $0$ was taken to generalize standard accessibility. That is, for every formula $\varphi$ of $\PDL$ we define a formula $\sigma(\varphi)$ of $\NPDL$ inductively by the rules 
\begin{itemize}
    \item $\sigma(p) = p$ 
    \item $\sigma(\neg \varphi) = \neg \sigma(\varphi)$
    \item $\sigma(\varphi_1 \wedge \varphi_2) = \sigma(\varphi_1) \wedge \sigma(\varphi_2)$
    \item $\sigma(\diam{\pi}\varphi) = \diam{\pi}^0\sigma(\varphi)$
\end{itemize}
Given a model $\mathfrak{N} = \langle X, \{R_a\}_{\AtProg}, \Val \rangle$ of $\PDL$ (\emph{i.e.}, $R_a \subseteq X^2$ are non-weighted atomic accessibility relations), we define a corresponding $\NPDL$-model $\mathcal{S}(\mathfrak{N}) = \langle X, \{R'_a\}_{\AtProg}, \Val \rangle$ by 
$$
R'_a (x,y)  = \begin{cases}
    0 & \text{ if } xR_a y, \\
    \infty & \text{ otherwise}.
\end{cases}
$$
By definition it is straightforward to check that the above extends to all programs $\pi$ and that 
$$
\mathfrak{N}, x \Vdash \varphi \quad \text{ if and only if } \quad \mathcal{S}(\mathfrak{N}), x \Vdash \sigma(\varphi).
$$ Since the translation $\sigma$ does not increase the length of the formula $\varphi$, we obtain our lower bound from the fact that $\PDL$ is \textsc{exptime}-complete (e.g.\ see \cite[Chapter 8]{HarelKozen2000}). 

\begin{corollary}[Lower Bound]\label{Corollary:ComplexityLowerBound}
Satisfiability of $\NPDL$ is \emph{\textsc{exptime}}-hard.  
\end{corollary}

In order to obtain an upper bound as well, we now present a translation $\tau$ of $\NPDL$ into standard $\PDL$ over the set of atomic programs 
$$
\AtProg^{=\N} := \{ a^{=m} \mid a \in \AtProg, m\in \mathbb{N} \}.
$$
The idea is that the standard $\PDL$ program $a^{=m}$ encodes the information $R_a(x,y) = m$ in a model of $\NPDL$. From this, one can recover complex $\NPDL$ program constructs as follows.

\begin{definition}[Formula translation]\label{definition:FormulaTranslationPDL}
By simultaneous induction, for all formulas $\varphi$ and programs $\pi$ of $\NPDL$ we define the formula $\tau(\varphi)$ and for every $m\in \N$ the program $\tau^{\leq m}(\varphi)$ of $\PDL$ (over the set of atomic programs $\AtProg^{=\N}$ as above) as follows.
\\
\textbf{\textit{Formulas}:}
\begin{itemize}
\item $\tau(p) = p$ for propositional variables  
\item $\tau(\neg \varphi) = \neg \tau(\varphi)$  
\item $\tau(\varphi_1 \wedge \varphi_2) = \tau(\varphi_1) \wedge \tau(\varphi_2)$ 
\item $\tau(\diam{\pi}^m \varphi) = \diam{\tau^{\leq m}(\pi)} \tau(\varphi)$
\end{itemize}
\textbf{\textit{Programs}:}
\begin{itemize}
\item $\tau^{\leq m}(a) = \bigcup_{i = 0}^{m} a^{= m}$ for atomic programs 
\item $\tau^{\leq m}(\pi_1 \cup \pi_2) = \tau^{\leq m}(\pi_1) \cup \tau^{\leq m}(\pi_2)$
\item $\tau^{\leq m}(\pi_1;\pi_2) = \bigcup_{i = 0}^{m} \tau^{\leq i}(\pi_1);\tau^{\leq m-i}(\pi_2)$
\item $\tau^{\leq 0}(\pi^\ast) = \tau^{\leq 0}(\pi)^\ast \phantom{i}$ and \\ $\tau^{\leq m}(\pi^\ast) = \tau^{\leq 0}(\pi^\ast) \cup \bigcup_{i = 1}^m \tau^{\leq 0}(\pi^\ast);\tau^{\leq i}(\pi);\tau^{\leq m-i}(\pi^\ast)$
\item $\tau^{\leq m}(\varphi?) = \tau(\varphi)?$
\end{itemize}
\end{definition}

As mentioned above, we now explain how to transform a model of $\NPDL$ into a model of $\PDL$ over the extended collection of atomic programs $\AtProg^{=\N}$. 

\begin{definition}[Model transformation]\label{definition:ModelTransformationPDL}
For $\mathfrak{M} = \langle X, \{ R_a \}_{\AtProg}, \Val \rangle$ a $\NPDL$-model, we define the $\PDL(\AtProg^{=\N})$ model 
$
\Tau(\mathfrak{M}) = \langle X, \{ R_{a^{=m}} \}_{\AtProg^{=\N}}, \Val \rangle
$
by $x R_{a^{=m}} y$ if and only if $R_a(x,y) = m$.  
\end{definition}

The main result of this subsection shows that the translation $\tau$ works as desired. 

\begin{theorem}[Translation Theorem]\label{theorem:TranslationTheorem}
Let $\mathfrak{M}$ and $\Tau(\mathfrak{M})$ be as in Definition~\ref{definition:ModelTransformationPDL}. Then for every $\NPDL$-formula $\varphi$ and state $x \in X$ it holds that 
$$
\mathfrak{M}, x \Vdash \varphi \quad \text{ if and only if } \quad \Tau(\mathfrak{M}), x \Vdash \tau(\varphi).
$$ 
\end{theorem}

\begin{proof}
By simultaneous induction, for every formula $\varphi$ we prove the above statement and for every program $\pi$ we show the following (for all $x,y \in X$): 
$$
R_{\pi}(x,y) \leq m \quad \text{ if and only if } \quad  x R_{\tau^{\leq m}(\pi)} y. 
$$ 
 
\textbf{Formula induction.} The base case $p \in \Prop$ is due to the fact that the same valuation $\Val$ is used in $\mathfrak{M}$ and $\Tau(\mathfrak{M})$. The cases $\neg \psi$ and $\psi_1 \wedge \psi_2$ are standard induction arguments.

\emph{$\bullet$ Case $\diam{\pi}^m \varphi$} : We have $\mathfrak{M}, x \Vdash \diam{\pi}^m\varphi$ if and only if there is $y \in X$ such that $R_\pi(x,y) \leq m$ and $\mathfrak{M}, y \Vdash \varphi$. By the inductive hypothesis on $\pi$ the former is equivalent to $x R_{\tau^{\leq m}(\pi)}y$ and by the inductive hypothesis on $\varphi$ the latter is equivalent to $\Tau(\mathfrak{M}), y \Vdash \tau(\varphi)$. Thus $\mathfrak{M},x \Vdash \varphi$ if and only if $\Tau(\mathfrak{M}), x \Vdash \diam{\tau^{\leq m}(\pi)}\tau(\varphi)$ as desired (recall $\tau(\diam{\pi}^m \varphi) = \diam{\tau^{\leq m}(\pi)} \tau(\varphi)$)  

\textbf{Program induction.} The base case $\pi=a  \in \AtProg$ is immediate from the definitions.

\emph{$\bullet$ Case $\pi_1 \cup \pi_2$} : We observe 
{\small
\begin{align*}
&x R_{\tau^{\leq m}(\pi_1 \cup \pi_2)} y  \phantom{i}\Longleftrightarrow\phantom{i} (x,y) \in R_{\tau^{\leq m}(\pi_1)} \cup R_{\tau^{\leq m}(\pi_2)} && \text{(Def. $\tau^{\leq m }(\pi_1 \cup \pi_2)$)}  \\
&\phantom{i}\Longleftrightarrow\phantom{i} R_{\pi_1}(x,y) \leq m \text{ or } R_{\pi_2}(x,y) \leq m && \text{(IH on $\pi_1$, $\pi_2$)} \\
&\phantom{i}\Longleftrightarrow\phantom{i} R_{\pi_1}(x,y) \wedge R_{\pi_2}(x,y) \leq m \phantom{i}\Longleftrightarrow\phantom{i} R_{\pi_1 \cup \pi_2}(x,y) \leq m. && \text{(Def. $R_{\pi_1 \cup \pi_2}$)}
\end{align*}
}

\emph{$\bullet$ Case $\pi_1 ; \pi_2$} : We observe
{\small 
\begin{align*}
 & x R_{\tau^{\leq m}(\pi_1 ; \pi_2)} y \phantom{i}\Longleftrightarrow\phantom{i} (x,y) \in  \bigcup_{i = 0}^{m} R_{\tau^{\leq i}(\pi_1)};R_{\tau^{\leq m-i}(\pi_2)} && \text{(Def. $\tau^{\leq m}(\pi_1 ; \pi_2)$)} \\
 & \phantom{i}\Longleftrightarrow\phantom{i} \bigvee_{i = 0}^{m} \exists v \colon (x,v) \in R_{\tau^{\leq i}(\pi_1)} \ \& \ (v,y) \in R_{\tau^{\leq m-i}(\pi_2)} && \text{(Def. $\PDL$-programs)}\\
 & \phantom{i}\Longleftrightarrow\phantom{i} \bigvee_{i = 0}^{m} \exists z \colon R_{\pi_1}(x,z) \leq i  \ \& \ R_{\pi_2}(z,y) \leq m-i && \text{(IH on $\pi_1$, $\pi_2$)} \\
 & \phantom{i}\Longleftrightarrow\phantom{i} \bigwedge_{v \in X} R_{\pi_1}(x,v) + R_{\pi_2}(v,y) \leq m  \phantom{i}\Longleftrightarrow\phantom{i} R_{\pi_1;\pi_2}(x,y) \leq m. && \text{(Def. $R_{\pi_1;\pi_2}$)}
\end{align*}
} 

\emph{$\bullet$ Case $\tau^{\leq m}(\pi^\ast)$} : By induction on $m$. For the base case $m = 0$ we observe
{\small 
\begin{align*}
 &x R_{\tau^{\leq 0}(\pi^\ast)} y  \phantom{i}\Longleftrightarrow\phantom{i} x R_{\tau^{\leq 0}(\pi)}^\ast y && \text{(Def. $\tau^{\leq 0}(\pi^\ast)$)} \\
 & \phantom{i}\Longleftrightarrow\phantom{i} x = y \text{ or } \exists v_1, \dots, v_k \colon xR_{\tau^{\leq 0}(\pi)}v_1 \dots v_kR_{\tau^{\leq 0}(\pi)}y  && \text{(Def. $\PDL$-iteration)} \\
 & \phantom{i}\Longleftrightarrow\phantom{i} x = y \text{ or } \exists v_1, \dots, v_k \colon R_{\pi}(x,v_1) \wedge \dots \wedge R_{\pi}(v_k,y) = 0 && \text{(IH on $\pi$)} \\ 
 & \phantom{i}\Longleftrightarrow\phantom{i} R_{\pi^\ast}(x,y) \leq 0. && \text{(Def. $R_{\pi^\ast}$)}
\end{align*}
}
For the inductive step, note that in the following we have $i \geq 1$, thus $m-i < m$. 
{\small 
\begin{align*}
 &x R_{\tau^{\leq 0}(\pi^\ast)} y  \phantom{i}\Longleftrightarrow\phantom{i} (x,y) \in R_{\tau^{\leq 0}(\pi)}^\ast \cup \bigcup_{1 \leq i \leq m} R_{\tau^{\leq 0}(\pi^\ast)};R_{\tau^{\leq i}(\pi)};R_{\tau^{\leq m-i}(\pi^\ast)}  \\
 & \phantom{i}\Longleftrightarrow\phantom{i} R_{\pi^\ast}(x,y) \leq 0 \text{ or } \bigvee_{1 \leq i \leq m} \exists v_1, v_2 \colon x R_{\tau^{\leq 0}(\pi^\ast)}v_1 R_{\tau^{\leq i}(\pi)} v_2 R_{\tau^{\leq m-i}(\pi^\ast)}y  \\
 & \phantom{i}\Longleftrightarrow\phantom{i} R_{\pi^\ast}(x,y) \leq 0 \text{ or } \bigvee_{1 \leq i \leq m} \exists v_1, v_2 \colon  R_{\pi^\ast}(x,v_1) \leq 0,  R_{\pi}(v_1,v_2) \leq i,  R_{\pi^\ast}(v_2,y) \leq m-i\\ 
 & \phantom{i}\Longleftrightarrow\phantom{i} R_{\pi^\ast}(x,y) \leq m. 
\end{align*}
}
The first `$\Leftrightarrow$' uses the definition of $\tau^{\leq m}(\pi^\ast)$, the second one uses the case $m = 0$ and definition of $\PDL$-programs, the third one uses the inductive hypothesis on $\pi$ and $m-i < m$.    

\emph{$\bullet$ Case $\tau^{\leq m}(\varphi?)$} : We observe
{\small
\begin{align*}
&R_{\varphi?}(x,y) \leq m  \phantom{i}\Longleftrightarrow\phantom{i} x = y \text{ and } \mathfrak{M},x \Vdash \varphi && \text{(Def. $R_{\varphi?}$)}  \\
&\phantom{i}\Longleftrightarrow\phantom{i} x = y \text{ and } \Tau(\mathfrak{M}), x \Vdash \tau(\varphi) && \text{(IH on $\varphi$)} \\
&\phantom{i}\Longleftrightarrow\phantom{i} x R_{\tau^{\leq m}(\varphi?)}y. && \text{(Def. $\tau^{\leq m}(\varphi?)$)}
\end{align*}
}
This finishes the proof.
\end{proof}

Recall that $N(\varphi)$ is the largest index of modalities occurring in $\varphi$ (Definition~\ref{definition:BoundOnWeights}), furthermore we use $|\varphi|$ to denote the length of $\varphi$ (brackets and indices of modalities excluded). The following establishes the growth in size of a formula under the translation $\tau$ in terms of these two parameters.  

\begin{lemma}\label{lemma:TranslationLength}
For every formula $\varphi$, the length of $\tau(\varphi)$ is $O(2^{2|\varphi| N(\varphi)}|\varphi|)$. Similarly, for every program $\pi$, the length of $\tau^{\leq m}(\pi)$ is $O(2^{|\pi|(m+N(\pi))}|\pi| )$.
\end{lemma}
\begin{proof}
Simultaneous induction on formula $\varphi$, program $\pi$ and index $m$. The base cases of $a \in \AtProg$ and $p \in \Prop$ are clear by definitions. 
\\
\textbf{\textit{Formula Induction}.} 

\emph{$\bullet$ Case $\neg \varphi$} :  
{\small
\begin{align*}
&|\tau (\neg\varphi)| = |\tau (\varphi)|+1 \\
& = O\big( (2^{2|\varphi| N(\varphi)})|\varphi|)+1\big) &&\text{(Inductive Hypothesis)} \\
& = O\big( (2^{2|\varphi| N(\varphi)})(|\varphi|+1)\big) \\
& = O\big( (2^{2|\neg \varphi| N(\neg \varphi)})|\neg \varphi|\big).
\end{align*}
}

\emph{$\bullet$ Case $\varphi_1 \wedge \varphi_2$} :
{\small
\begin{align*}
&|\tau (\varphi_1 \wedge \varphi_2)|  =  |\tau (\varphi_1)| +|(\tau (\varphi_2)| +1 \\
& = O\big( 2^{2|\varphi_1|N(\varphi_1)}|\varphi_1|  +2^{2|\varphi_2| N(\varphi_2)})|\varphi_2| +1\big) &&  \text{(Inductive Hypothesis)} \\
& = O\big( (2^{2(|\varphi_1|+|\varphi_2|)N(\varphi_1 \wedge \varphi_2)})(|\varphi_1|+|\varphi_2|) +1)\big) \\ 
& = O\big( (2^{2|\varphi_1 \wedge \varphi_2|N(\varphi_1 \wedge \varphi_2)})|\varphi_1\wedge \varphi_2|\big).
\end{align*}
}
\emph{$\bullet$ Case $\diam{\pi}^m(\varphi)$} :
{\small
\begin{align*}
&|\tau (\diam{\pi}^m\varphi)|  = |\tau^{\leq m}(\pi)|+ |\tau (\varphi)| \\
& = O\big(2^{2|\varphi| N(\varphi)}|\varphi|  +2^{|\pi| (m+N(\pi))})|\pi|\big) && \text{(Inductive Hypotheses)} \\
& = O\big((2^{2|\varphi| N(\diam{\pi}^m \varphi)}|\varphi|  +2^{2|\pi|N(\diam{\pi}^m \varphi) })|\pi|\big) && \text{($m \leq N(\diam{\pi}^m \varphi)$)} \\
& = O\big(2^{2|\diam{\pi}^m \varphi| N(\diam{\pi}^m \varphi)}|\diam{\pi}^m \varphi|\big).
\end{align*}
}
\textbf{\textit{Program Induction}.} 

\emph{$\bullet$ Case $\pi_1 \cup \pi_2$} :
{\small
\begin{align*}
&|\tau^{\leq m} (\pi_1 \cup \pi_2)|  = | \tau^{\leq m} (\pi_1)| + | \tau^{\leq m} (\pi_2)|+1 \\
& = O\big((2^{|\pi_1| (m+N(\pi_1))}|\pi_1|+2^{|\pi_2| (m+N(\pi_2))})|\pi_2|) + 1\big) && \text{(Inductive Hypothesis)} \\
& = O\big(2^{|\pi_1 \cup \pi_2|(m+N(\pi_1 \cup \pi_2) )}(|\pi_1|+|\pi_2|+1)\big) \\ 
& = O\big(2^{|\pi_1 \cup \pi_2|(m+N(\pi_1 \cup \pi_2)) )}|\pi_1 \cup \pi_2|\big).
\end{align*}
}

\emph{$\bullet$ Case $\pi_1; \pi_2$} :
{\small
\begin{align*}
&|\tau^{\leq m} (\pi_1 ; \pi_2)|  = |\bigcup_{i = 0}^{m} \tau^{\leq i}(\pi_1);\tau^{\leq m-i}(\pi_2)| \\
& = \sum_{i=0}^m (  | \tau^{\leq i} (\pi_1)|+ | \tau^{\leq m-i} (\pi_2)|+2) \\ 
& = O\big(\sum_{i=0}^m (2^{|\pi_1| (i+N(\pi_1))}|\pi_1|+2^{|\pi_2| (m-i+N(\pi_2))})|\pi_2|+2)\big) && \text{(Inductive Hypothesis)} \\ 
& = O\big( (2^{|\pi_1| (m+N(\pi_1))+1}|\pi_1|+2^{|\pi_2| (m+N(\pi_2))+1})|\pi_2|+2m)\big) && \text{($\sum_{i=1}^k 2^i \leq 2^{k+1}$)} \\ 
& = O\big( 2^{(|\pi_1|+|\pi_2|+1)(m+N(\pi_1 ; \pi_2))}(|\pi_1|+|\pi_2|+1)\big) \\ 
& = O\big( 2^{|\pi_1; \pi_2|(m+N(\pi_1 ; \pi_2))}|\pi_1; \pi_2|\big).
\end{align*}
}

\emph{$\bullet$ Case $\pi^\ast$} : For $m=0$, observe
{\small
\begin{align*}
&|\tau^{\leq 0} (\pi^\ast)|  = |\tau^{\leq 0} (\pi)^\ast | \\ 
& = O(2^{|\pi|(N(\pi))}|\pi| ) +1 && \text{(Inductive Hypothesis)} \\
& = O\big(2^{|\pi|(N(\pi))}(|\pi|+1)\big) \\ 
& = O\big(2^{|\pi^\ast|(N(\pi^\ast))}|\pi^\ast|\big). 
\end{align*}
}
\noindent For $m\geq 1$, observe
{\small
\begin{align*}
&|\tau^{\leq m} (\pi^\ast)|  = |\tau^{\leq 0} (\pi)^\ast) \cup \bigcup_{i = 1}^m \tau^{\leq 0}(\pi^\ast);\tau^{\leq i}(\pi);\tau^{\leq m-i}(\pi^\ast)| \\
& = |\tau^{\leq 0} (\pi)^\ast| +\sum_{i = 1}^m (|\tau^{\leq 0}(\pi^\ast)|+ |\tau^{\leq i}(\pi)|+ |\tau^{\leq m-i}(\pi^\ast)| +3) \\
& = O\big(2^{|\pi^\ast|N(\pi^\ast)}|\pi^\ast| +\sum_{i = 1}^m(2^{|\pi^\ast|N(\pi^\ast)}|\pi^\ast| +\\ 
&\phantom{=}+2^{|\pi|(i+N(\pi))}|\pi|  + 2^{|\pi^\ast|(m-i+N(\pi^\ast))}|\pi^\ast|+3)\big) && \text{(Inductive Hypothesis)} \\ 
& = O\big(2^{|\pi^\ast|M(\pi^\ast)}|\pi^\ast| +m2^{|\pi^\ast|M(\pi^\ast)}|\pi^\ast| + \\
&\phantom{=}+2^{|\pi|(m+N(\pi))+1}|\pi| + 2^{|\pi^\ast|(m+N(\pi^\ast))}|\pi^\ast|+3m) && \text{($\sum_{i=1}^k 2^i \leq 2^{k+1}$)} \\
& = O\big(2^{|\pi^\ast|(m+N(\pi^\ast))}|\pi^\ast|\big).
\end{align*}
}
The last equality holds since $m <2^m$ and $|\pi|(m+N(\pi))+1 < |\pi^\ast|(m+N(\pi^\ast))$, asymptotically the term $2^{2|\pi|(m+M(\pi))+1}|\pi|$ dominates all other terms in the preceding sum. 

\emph{$\bullet$ Case $\varphi?$} : We assume that in any formula $\varphi$, all occurrences of $\diam{\varphi?}^m$ are replaced by $\diam{\varphi?}^{N(\varphi)}$. The formula obtained after this substitution is equivalent to the original one. 
{\small
\begin{align*}
&|\tau^{\leq m} (\varphi ?)|  = | \tau(\varphi)|+1 \\ 
&  =O\big( 2^{2|\varphi| N(\varphi)}|\varphi|+1\big) && \text{(Inductive Hypothesis)} \\
& = O\big( (2^{|\varphi| (m+N(\varphi?))})(|\varphi|+1)\big) \\ 
& = O\big( 2^{|\varphi?|(m+N(\varphi?))}|\varphi?|\big) && \text{($m=N(\varphi)$ by assumption)}
\end{align*}
}
This finishes the proof.    
\end{proof}

Note that this also indicates that $\NPDL$ is more succinct than $\PDL$. Combining the above lemma this with the fact that $\PDL$ is \textsc{exptime}-complete, we obtain our upper complexity bound.  

\begin{corollary}[Upper Bound]\label{Corollary:ComplexityUpperBound}
The satisfiability problem of $\NPDL$ is in \emph{\textsc{2exptime}}.  
\end{corollary}

\begin{remark}[Adding converse and intersection]\label{remark:ComplexityConverseIntersection}
The translation $\tau$ can be extended to converse programs $\pi^\smallsmile$ interpreted as $R_{\pi^\smallsmile}(x,y) = R_{\pi}(y,x)$ via $\tau^{\leq m}(\pi^\smallsmile) = \tau^{\leq m}(\pi)^\smallsmile$, hence we obtain the same complexity bounds for this extension. Similarly, $\tau$ can be extended to \emph{concurrent} $\NPDL$ \cite{Danecki1985} with intersection generalized by $R_{\pi_1 \cap \pi_2}(x,y) = R_{\pi_1}(x,y) \vee R_{\pi_2}(x,y)$ via $\tau^{\leq m}(\pi_1 \cap \pi_2) = \tau^{\leq m}(\pi_1 ) \cap \tau^{\leq m}(\pi_2)$. Hence we obtain decidability of concurrent $\NPDL$ and from \textsc{2exptime}-completeness of concurrent $\PDL$ \cite{Mayer1996} we obtain a \textsc{3exptime}-upper bound.
\end{remark}

\subsection{The Model Checking Problem}\label{subsection:ModelCheckingComplexity}
In this subsection, we show that given a $\NPDL$ model $\mathfrak{M} = (X, R_a, \Val)$, a state $x \in X$, and a formula $\varphi$,  we can check whether $\mathfrak{M},x \models \varphi$ in time $O(|\mathfrak{M}|^3|\varphi|)$.
The model checking is performed by computing the set of worlds in $\mathfrak{M}$ which satisfy $\varphi$  (denoted by $\sem{\varphi}$) through the following recursive steps on the complexity of $\varphi$.
\begin{enumerate}
    \item $\sem{p}:=\Val(p)$.
    \item $\sem{\varphi_1 \wedge \varphi_2}:=\sem{\varphi_1} \cap \sem{\varphi_2}$.
    \item $\sem{\neg\varphi}:=X{\setminus}\sem{\varphi}$.
    \item  $\sem{\diam{\pi}^m \varphi}:= {\pi^{\leq m}}^{-1}(\sem{\varphi}) = \{x \mid \exists y \in  \sem{\varphi} \colon  (R_\pi(x, y) \leq m)\}$.
\end{enumerate}
For any $x, y \in X$, and any program $\pi$, given $R_a(x,y)$ for any atomic program $a$ occurring in $\pi$, we compute $R_\pi(x,y)$ by the following recursion. 
\begin{enumerate}
    \item $R_{\pi_1 \cup \pi_2}(x,y) = \min (R_{\pi_1}(x,y),  R_{\pi_2}(x,y))$. 
    \item $R_{\pi_1 \circ \pi_2}(x,y) = \min_{v \in X} R_{\pi_1} (x,v)+ R_{\pi_2} (v,y)$.
    \item For any $\pi$, given $R_\pi$, we can compute $R_{\pi^*}$ in time $O(|\mathfrak{M}|^3)$ using the Floyd–Warshall algorithm \cite{floyd1962algorithm,10.1145/321105.321107}. 
\end{enumerate}

\begin{proposition}\label{proposition:ComputeModel}
Let $\pi$ be a program and $\varphi$ be a formula.
\begin{enumerate}
    \item The relation $R_\pi$ can be computed in time $O(|\mathfrak{M}|^3|\pi|)$.
    \item The set $[\![\varphi]\!]$ can be computed in time $O(|\mathfrak{M}|^3|\varphi|)$.
\end{enumerate}
\end{proposition}
\begin{proof}
By simultaneous induction we show \emph{(1)} for every program $\pi$ and \emph{(2)} for every formula $\varphi$. 
\\
\textbf{\textit{Program Induction}.} 
The base case is trivially true. 

$\bullet$ \emph{Case} $\pi_1 \cup \pi_2$ : If $R_{\pi_1}$ and  $R_{\pi_2}$ are already computed, we need to perform one new operation (taking minimum) for any pair of points $x, y \in X$. Therefore (using the inductive hypothesis), the total number of steps in this computation is 
bounded by $O(|\mathfrak{M}|^3|\pi_1|)+ O(|\mathfrak{M}|^3|\pi_2|)+O(|\mathfrak{M}|^2|=O(|\mathfrak{M}|^3|\pi_1 \cup \pi_2|) $. 

$\bullet$ \emph{Case} $\pi_1; \pi_2 $ : If $R_{\pi_1}$ and  $R_{\pi_2}$ are already computed, we need to perform two new operations (taking sum and minimum) for any 
triplet  of points $x, v, y \in W$. Therefore (by the inductive hypothesis), the total number of steps in the computation is 
bounded by $O(|\mathfrak{M}|^3|\pi_1|)+ O(|\mathfrak{M}|^3|\pi_2|)+O(|\mathfrak{M}|^3|=O(|\mathfrak{M}|^3|\pi_1 \circ \pi_2|) $.

$\bullet$ \emph{Case} $\pi^* $ : We need to do  $O(|\mathfrak{M}|^3)$ new operations using the Floyd-Warshall algorithm. Therefore, by induction, the total number of operations is bounded by $O(|\mathfrak{M}|^3|\pi|)+ O(|\mathfrak{M}|^3)=O(|\mathfrak{M}|^3|\pi^*)$. 

$\bullet$ \emph{Case} $\pi=\varphi?$ : By definition, $R_{\varphi?}= [\![\varphi]\!] \times [\![\varphi]\!]$. Therefore by induction, the total number of operations is bounded by  $O(|\mathfrak{M}|^3|\varphi|) +O(|\mathfrak{M}|^2) = O(|\mathfrak{M}|^3|\varphi?|)$. 
\\
\textbf{\textit{Formula Induction}.} 
The base case is trivially true. 

$\bullet$ \emph{Case} $\varphi = \varphi_1 \wedge \varphi_2 $ :  
Since  $[\![\varphi]\!]=[\![\varphi_1]\!] \cap [\![\varphi_2]\!]$, we need to do $O(|\mathfrak{M}|)$ new operations.
Therefore, the total number of operations is bounded by  $O(|\mathfrak{M}|^3|\varphi_1|) + O(|\mathfrak{M}|^3|\varphi_2|) +O(|\mathfrak{M}|) = O(|\mathfrak{M}|^3|\varphi_1 \wedge \varphi_2|)$. 

$\bullet$ \emph{Case} $\varphi=\neg \varphi_1 $ : 
Since  $[\![\varphi]\!]=X {\setminus} [\![\varphi_1]\!]$, we need to do $O(|\mathfrak{M}|)$ new operations.
Therefore, the total number of operations is bounded by  $O(|\mathfrak{M}|^3|\varphi_1|) + O(|\mathfrak{M}|) = O(|\mathfrak{M}|^3|\neg \varphi_1|)$.

$\bullet$ \emph{Case} $\varphi=\diam{\pi}^m \varphi_1 $ : 
since  $[\![\varphi]\!]= \{x \mid \exists y \in  \sem{\varphi_1} (R_\pi(x, y) \leq m)\}$, we need to do $O(|\mathfrak{M}|^2)$ new operations.
Therefore, the total number of operations is bounded by  $O(|\mathfrak{M}|^3|\varphi_1|) +O(|\mathfrak{M}|^3|\pi|) +O(|\mathfrak{M}|^2) = O(|\mathfrak{M}|^3| \diam{\pi}^m\varphi_1|)$. 
\end{proof}

In conclusion, we obtain the following complexity bound on the model checking problem. Note that this bound is a significant improvement over the naive bound obtained by the translation of $\NPDL$ into $\PDL$ of the previous subsection.    

\begin{corollary}\label{corollary:ModelCheckingComplexity}
The model checking problem of $\NPDL$ is in $O(|\mathfrak{M}|^3|\varphi|)$.   
\end{corollary}

\section{Conclusion and Future Research}\label{section:Conclusion}
We introduced $\NPDL$, a variant of propositional dynamic logic interpreted on $\mathbb{N}^\infty$-weighted frames with modalities $\diam{\pi}^m$ describing accessibility with resource-consumption at most $m \in \mathbb{N}$. Showing that satisfiability can be checked on finite models of bounded size and with bounded weights, we proved that $\NPDL$ is decidable. Combining filtration techniques with the canonical model construction, we showed completeness. Lastly, using translations into standard $\PDL$ we obtained complexity bounds for $\NPDL$. 

In this work, we have provided a \textsc{2exptime}-upper bound on the \emph{complexity} of $\NPDL$. The main open question at this point is whether that bound is strict, or whether $\NPDL$ remains \textsc{exptime}-complete (e.g.\ for the case of $\PDL$, see Pratt's algorithm \cite{Pratt1980}). This problem also remains open for extensions of $\NPDL$ with additional program constructs such as converse and intersection (recall Remark~\ref{remark:ComplexityConverseIntersection}).

Another question is whether $\NPDL$ has \emph{Craig Interpolation}, a property which has only recently been shown for standard $\PDL$ \cite{Borzechowski2025} and converse $\PDL$ \cite{Kloibhofer2026}. Related to this, in future work we aim to develop the \emph{proof theory} of $\NPDL$, for instance via cyclic tableaux systems where linear inequalities of weights are explicitly included in the syntax.  

For further future research directions we propose to consider cost-sensitive variants of extensions and generalizations of $\PDL$, for instance of \emph{Game Logic} \cite{Parikh1983, Pauly2003} or (first-order) \emph{Dynamic Logic} \cite[Chapter 11]{HarelKozen2000}. One may also consider other quantales than $\mathbb{N}^\infty$ to represent resources, making it more challenging to come up with a finitary axiomatization (note that our reduction axioms depend on the fact that there are only finitely many ways to write $n$ as a binary sum). In addition, it would be interesting to investigate the relation of $\NPDL$ to many-valued versions of $\PDL$ previously proposed for modeling program costs \cite{Behounek2008,Sedlar2020,Sedlar2021} or to graded-belief based programs \cite{Laverny2005,Laverny2005b}. 
Lastly, let us mention that for such an analysis the \emph{coalgebraic perspective} \cite{HansenKupke2015,HansenPoiger2025} (also see Remark~\ref{remark:Coalgebra}) is promising.

\section*{Acknowledgments}
Funded by the European Union under the Operational Programme Jan Amos Komenský project no. {CZ.02.01.01/00/23\_025/0008724}. The second author was furthermore supported by the European Research Executive Agency ERA Fellowship project no. 101334342. 

\bibliographystyle{acm} 
\bibliography{References1}

\end{document}